\documentclass[twocolumn,showpacs,amsfonts,aps,prd,floatfix,nofootinbib,float,superscriptaddress]{revtex4-1}

\usepackage{color}
\usepackage{amsmath}
\usepackage{amssymb}
\usepackage{multirow}
\usepackage{graphicx}
\usepackage[utf8]{inputenc}

\newcommand{\avg}[1]{\langle #1 \rangle}
\let\oldRe\Re
\renewcommand{\Re}{\oldRe\;}

\newcommand{\dd}{\mathrm{d}}

\newcommand{\br}[1]{\left(#1 \right)}

\newcommand{\bc}[1]{\left\{#1 \right\}}

\newcommand{\mat}[1]{\:\mathbf{#1}\,}

\newcommand{\Obs}{\mathcal{O}}

\begin{document}

\title{Thermal dilepton rates and electrical conductivity of the QGP from the lattice}

\author{Heng-Tong Ding}
\email{hengtong.ding@mail.ccnu.edu.cn}
\affiliation{Key Laboratory of Quark \& Lepton Physics (MOE) and Institute of
  Particle Physics, Central China Normal University, Wuhan 430079, China}
\author{Olaf Kaczmarek}
\email{okacz@physik.uni-bielefeld.de}
\author{Florian Meyer}
\email{fmeyer@physik.uni-bielefeld.de}
\affiliation{Fakult\"at f\"ur Physik, Universit\"at Bielefeld,
33615 Bielefeld, Germany}
\date{\today}
\begin{abstract}
	We investigate the temperature dependence of the thermal dilepton 
	rate and the electrical conductivity of the gluon plasma 
	at temperatures of $1.1T_c$, $1.3T_c$ and $1.5T_c$
	in quenched QCD. Making use of
	non-perturbatively clover-improved Wilson valence quarks allows for a clean
	extrapolation of the vector meson correlation function to the continuum limit. We found that the vector 
	correlation function divided by $T^3$ is almost temperature independent in the current 
	temperature window. 
	The spectral functions are obtained by $\chi^2$ fitting 
        of phenomenologically inspired Ans\"atze for the spectral function
        to the continuum
	extrapolated correlator data, where the correlations between the data points
	have been included.  Systematic uncertainties arising
        from varying the Ans\"atze motivated from strong
	coupling theory as well as perturbation theory are discussed and estimated.
        We 
	found that the
	electrical conductivity of the hot medium, related to the slope of the
	vector spectral function at zero frequency and momentum, is
	$0.2C_{em}\lesssim \sigma/T\lesssim0.7C_{em}$ for $T=1.1T_c$ 
	and $0.2C_{em}\lesssim \sigma/T\lesssim0.4C_{em}$ for the higher
	temperatures.  The dilepton rates and soft photon rates,
	resulting from the obtained spectral functions, show no significant
	temperature dependence, either.
        
\end{abstract}

\maketitle

\section{Introduction} 
Ongoing Heavy Ion Collision experiments conducted at facilities like RHIC and
LHC provide new output about the nature of elementary particles and their
interactions. Direct photons and dileptons ($e^{+}e^{-},\mu^{+}\mu^{-}$) 
are especially good probes of the QGP,
as they are produced in all stages of its evolution and, 
since they are objects underlying 
the electroweak interaction, 
their coupling to QGP constituents is small \cite{David2006, Akiba2015}; once they are 
produced, they leave the interaction region largely unmodified. 
The latest experiments performed at PHENIX and STAR 
provide indications of thermal enhancements of dilepton spectra in 
the small to medium frequency region \cite{Adare2015,Adamcyk2015}, which indicates that
modifications 
by the surrounding thermal medium take place. On the other hand,
the spectral function in the vector channel at finite temperature provides
theoretical information on the thermal dilepton rates accessible in these experiments 
\cite{Rapp2009,Bernecker2011}, which renders it a worthwhile object to study from 
theory. Especially the small
frequency region of the spectral function contains information on important dynamical
quantities like the flavor diffusion constant and the electrical conductivity of the 
plasma \cite{Meyer2011,Hong2010}.
Because this regime is also inherently non-perturbative, 
the use of lattice QCD data is needed.
We will in the following attempt 
to determine the vector channel spectral function for light quark flavors from 
the theory of QCD.
With this we extend our former investigations 
\cite{Ding2010,Francis2012,Kaczmarek2013,Ding2014}. 
For other lattice QCD studies, based on finite lattices, see
\cite{Gupta:2003zh,Aarts2007,Brandt:2015aqk,Aarts:2014nba}.
Other determinations of the electrical conductivity can be found in 
\cite{Bandyopadhyay:2015wua,Linnyk:2015rco,Greif:2016skc}.\\

A well accessible quantity on the lattice is the correlation function
in a given mesonic channel.  It inhibits dynamical properties of the QGP when 
investigated at finite temperature. As such, the light vector
correlator is related to the electrical conductivity $\sigma$ of the QGP,
the dilepton rate $\frac{\dd W}{\dd \omega\dd^3p}$ and the photon rate
$\frac{\dd R}{\dd^3p}$
as accessible in heavy ion collision 
experiments, via the vector channel spectral function 
$\rho_V$ \cite{McLarren1984,Moore2006}.
While in general spectral functions relate to correlators through an integral 
equation,
\begin{eqnarray}
	\label{eqn_integ_trans}
	G(\tau,\vec{p})&=& \int\limits_0^{\infty}
	\!\frac{\dd \omega}{2\pi}\rho_H(\omega,\vec{p},T) K(\omega,\tau,T) \ \
        \ \ \\
	\text{with} \ &&
	K(\omega,\tau,T) = \frac{\cosh(\omega( \tau-\frac{1}{2T}))}{\sinh(\frac{\omega}{2T})},
\end{eqnarray}
transport coefficients are related to the spectral functions via Kubo formulas.
Examples of these are the shear and bulk viscosity obtained from energy momentum tensor
correlation functions, the heavy quark momentum diffusion coefficient from color electric 
correlators \cite{Kaczmarek2014,Francis2015II}, 
and the electrical conductivity, related to the light vector spectral function.
In the latter case, the Kubo formula is explicitly written as 
\begin{eqnarray}
	\label{eqn_kubo}
	\frac{\sigma}{T}=\frac{C_{em}}{6}\lim\limits_{\omega\rightarrow 0}
	\frac{\rho_{ii}(\omega,\vec{p}=\vec{0},T)}{\omega T},
\end{eqnarray}
with $\rho_{ii}$ denoting only the spatial components of the vector channel
and $C_{em} = e^2\sum q_f^2$ is the sum of the square of the individual quark charges.
The two experimentally observable rates mentioned above, originating from 
processes
at all stages of the collisions, can be written in terms of the 
spectral function in the vector channel and to leading order read
\cite{McLarren1984,Moore2006}
\begin{eqnarray} 
	\label{eqn_dilrate}
	\frac{\dd W}{\dd \omega
	\dd^3p} &=& \frac{C_{em}\alpha_{em}^2\rho_V(\omega,\vec{p},T)}
		{6\pi^3(\omega^2-\vec{p}^2)(e^{\omega/T}-1)},\\
	\label{eqn_photonrate}
		\omega \frac{\dd R_{\gamma}} {\dd^3p}&=&\frac{C_{em}\alpha_{em}\rho^T
	(\omega=|\vec{p}|,T)} {4\pi^2(e^{\omega/T}-1)}\;,
\end{eqnarray}
where $\rho_{V}$ is again the vector channel spectral function and 
$\rho^T$ is the spectral function transversally polarized with 
respect to the direction of $\vec{p}$.
These relations imply that once the spectral function of the vector 
channel is extracted from QCD, important insights into non-perturbative phenomena
of heavy ion collisions and the QGP can be gained.\\
In order to determine the spectral function, however, the Fredholm Type-I equation
(\ref{eqn_integ_trans}) has to be inverted,
which is often referred to as an "ill-posed" problem \cite{Meyer2011}. 
In our case it is a discrete problem, as we can access the value of the 
correlation function only at a finite number of points in $\tau T$.
The basic fact is that the numerical (temporal) correlator data contains
$\mathcal{O}(10)$ points, while a solution should be much more fine grained,
ideally even continuous. This means that there is more information 
desired on the r.h.s. than is actually provided on the l.h.s. of equation 
(\ref{eqn_integ_trans}). This simple lack 
of information identifies problem (\ref{eqn_integ_trans}) to be 
principally \textit{under-determined}.
Moreover, the available correlator 
data points are not known exactly either, as they stem from a Monte Carlo 
simulation and are subject to statistical uncertainties. This adds to the  
problem of having only a finite amount of input data. It can be investigated 
in some cases where it is 
possible to make a quantitative statement about the connection of fluctuations 
in the input data and fluctuations in the resulting solution, revealing a strong 
dependence of the solution on the accuracy of the input data.
This has been done in \cite{Bertero1985},
specifically for the Laplace kernel, or \cite{Hansen2007}, where essentially arbitrary integration
kernels are considered specifically within the framework of Tikhonov regularization. 
In addition to the finiteness of the data and their statistical uncertainties, 
within the framework of lattice QCD an important source of systematic errors are 
cutoff effects, arising from the lattice regularization itself, and that have 
to be removed by an extrapolation to the physical continuum.\\
\begin{table}[thbp]
	\centering
	\begin{tabular}{|c|c|c|c|c|c|c|c|c|c}
		\hline 
		${N_{\tau}}$ & ${N_{\sigma}}$ & ${\beta}$ &
		${\kappa}$ 
                & ${T\sqrt{t_0}}$ & ${T/T_c |^{ }_{t_0}}$ 
                & ${T r_0}$ & ${T/T_c |^{ }_{r_0}}$ 
                & $\mathrm{confs}$ \\ \hline \hline
		
		$32$ & $96$  & $7.192$ & $0.13440$ & $0.2796$ & $1.12$ & $0.8164$ & $1.09$ & $314$ \\ 
	        $48$ & $144$ & $7.544$ & $0.13383$ & $0.2843$ & $1.14$ & $0.8169$ & $1.10$ & $358$ \\ 
		$64$ & $192$ & $7.793$ & $0.13345$ & $0.2862$ & $1.15$ & $0.8127$ & $1.09$ & $242$  \\ \hline
		$28$ & $96$  & $7.192$ & $0.13440$ & $0.3195$ & $1.28$ & $0.9330$ & $1.25$ & $232$ \\ 
		$42$ & $144$ & $7.544$ & $0.13383$ & $0.3249$ & $1.31$ & $0.9336$ & $1.25$ & $417$ \\ 
		$56$ & $192$ & $7.793$ & $0.13345$ & $0.3271$ & $1.31$ & $0.9288$ & $1.25$ & $273$ \\ \hline
		$24$ & $128$ & $7.192$ & $0.13440$ & $0.3728$ & $1.50$ & $1.0886$ & $1.46$ & $340$  \\
	        $32$ & $128$ & $7.457$ & $0.13390$ & $0.3846$ & $1.55$ & $1.1093$ & $1.49$ & $255$ \\ 
		$48$ & $128$ & $7.793$ & $0.13340$ & $0.3817$ & $1.53$ & $1.0836$ & $1.45$ & $456$ \\ \hline
	\end{tabular}
	\caption{Parameters of all lattices used in this study.
        The $t_0$-scale \cite{Luscher:2010iy} and $r_0$-scale \cite{Sommer:1993ce} 
        and the conversion to $T_c$ are based
        on \cite{Francis2015}.
        Note that for the 
	continuum extrapolated data we state temperatures 
	of $T=1.1T_c$, $T=1.3T_c$ and $T=1.5T_c$ for the three data sets, respectively. 
        }
	\label{tab_lattice_data}
\end{table}
Any approach to solving an ill-posed problem, i.e. regain uniqueness and stability, must add 
information in order to "regularize" the problem and thus render it at least 
"better posed", with two important methods being the already mentioned Tikhonov regularization 
and the Maximum Entropy Method (MEM) \cite{Tarantola2004, Asakawa2000}
or related Bayesian methods \cite{Burnier2013}. Stochastic approaches
for the reconstruction of spectral functions
are currently under investigation \cite{Shu2015,Ohno2015}.
In a recent study \cite{Brandt:2015aqk} the Backus-Gilbert method was applied
for the determination of vector spectral functions and the electrical conductivity.\\
In the present work we choose the necessary additional information to 
enter the procedure in the form of a 
phenomenologically inspired Ansatz,
which is fitted to continuum extrapolated lattice QCD correlation functions.
Fixing the shape of the solution by supplying an Ansatz with two or 
three degrees of freedom is a very strong assumption, and the 
method of least 
squares fitting consequently is the natural tool to be employed. Because in this 
sense the problem has turned into an \textit{over-determined} one,
the choice of the Ansatz plays an essential role, and will be assessed 
critically by using different functional shapes in the fit.\\
The paper is organized as follows. 
After discussing the setup of the lattice calculation 
function and the continuum extrapolation of the vector correlation function in
the next section, we will introduce our Ansatz for the spectral function in
sec.~III and
discuss its properties and the thermal moments derived from it   
and analyze the the statitical uncertainties of the continuum extrapolated
correlators with a special emphasis on the importance of coveriances included
in our study.
Based on this, in sec.~IV we use a class of spectral functions fitted to the
continuum correlators to elaborate the systematic uncertainties for the
extraction of the vector spectral function. In sec.~V we present our final
results for the electrical conductivity, dilepton reates as well as soft photon
rates for the three temperatures used in this work, and conclude in sec.~VI.
\begin{figure*}[t]
	\centering
	\includegraphics[width=0.495\textwidth]{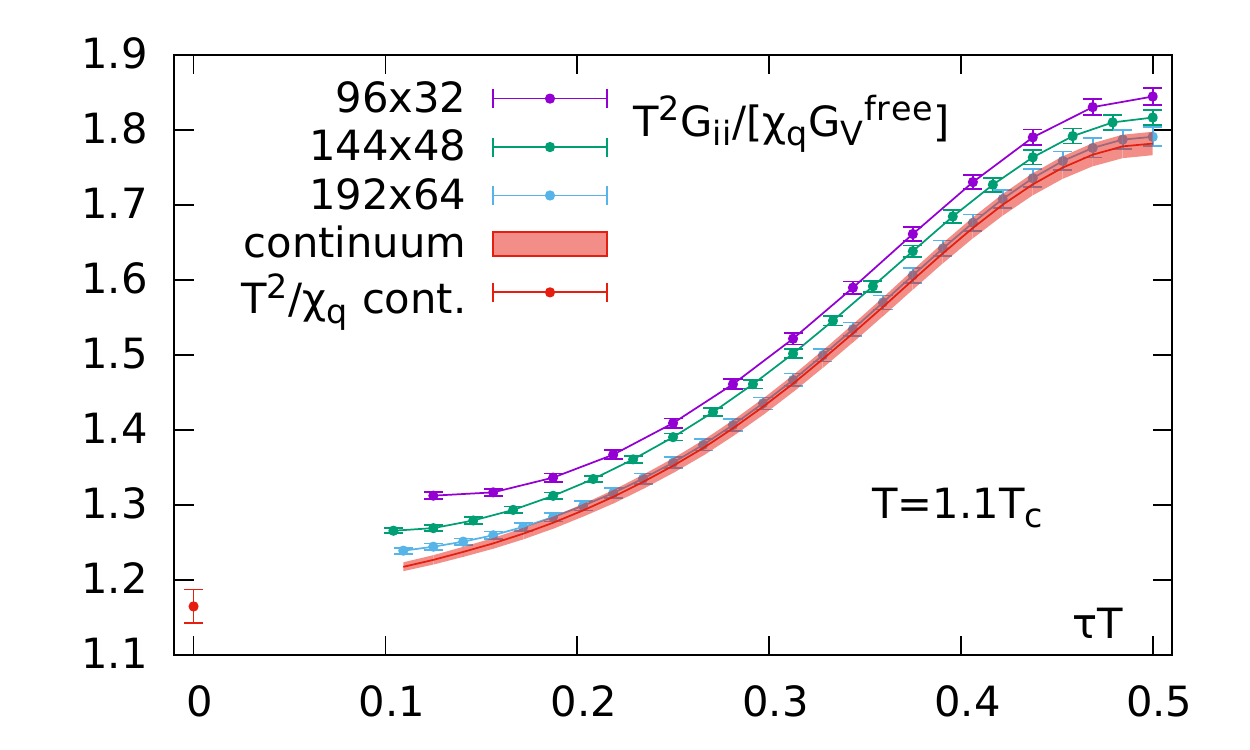}
	\includegraphics[width=0.495\textwidth]{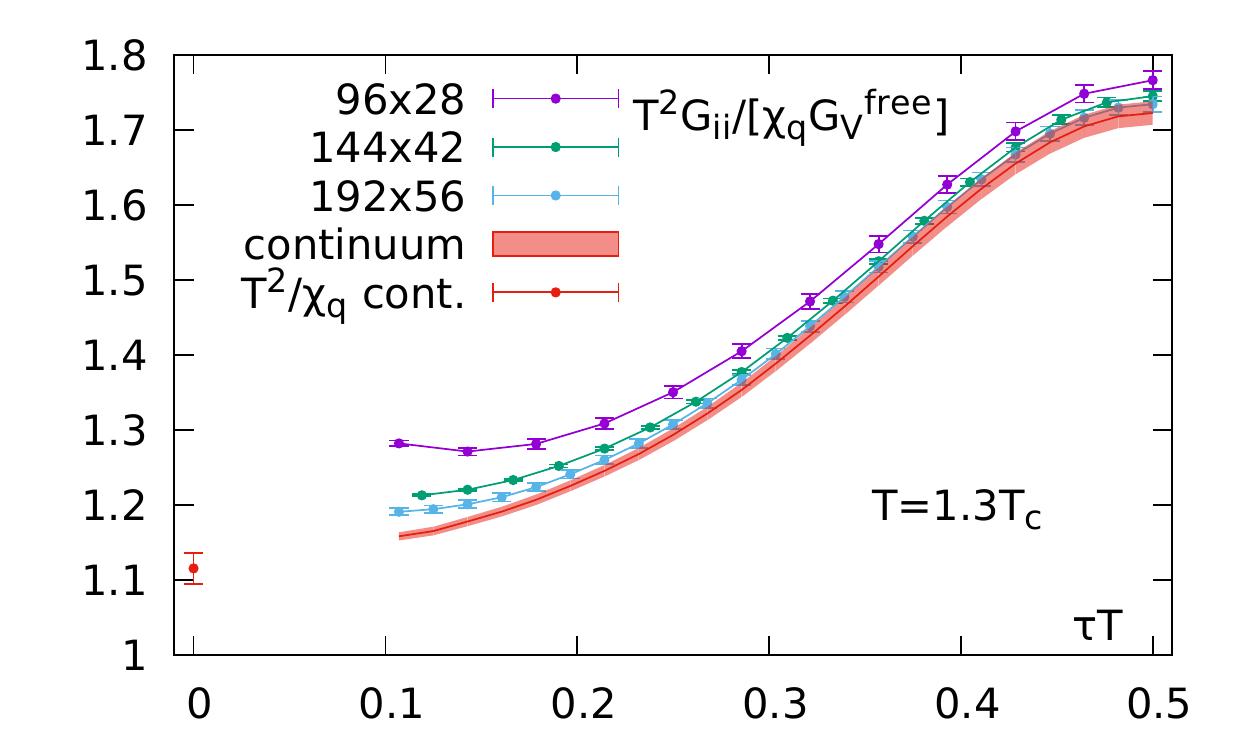}
	\includegraphics[width=0.495\textwidth]{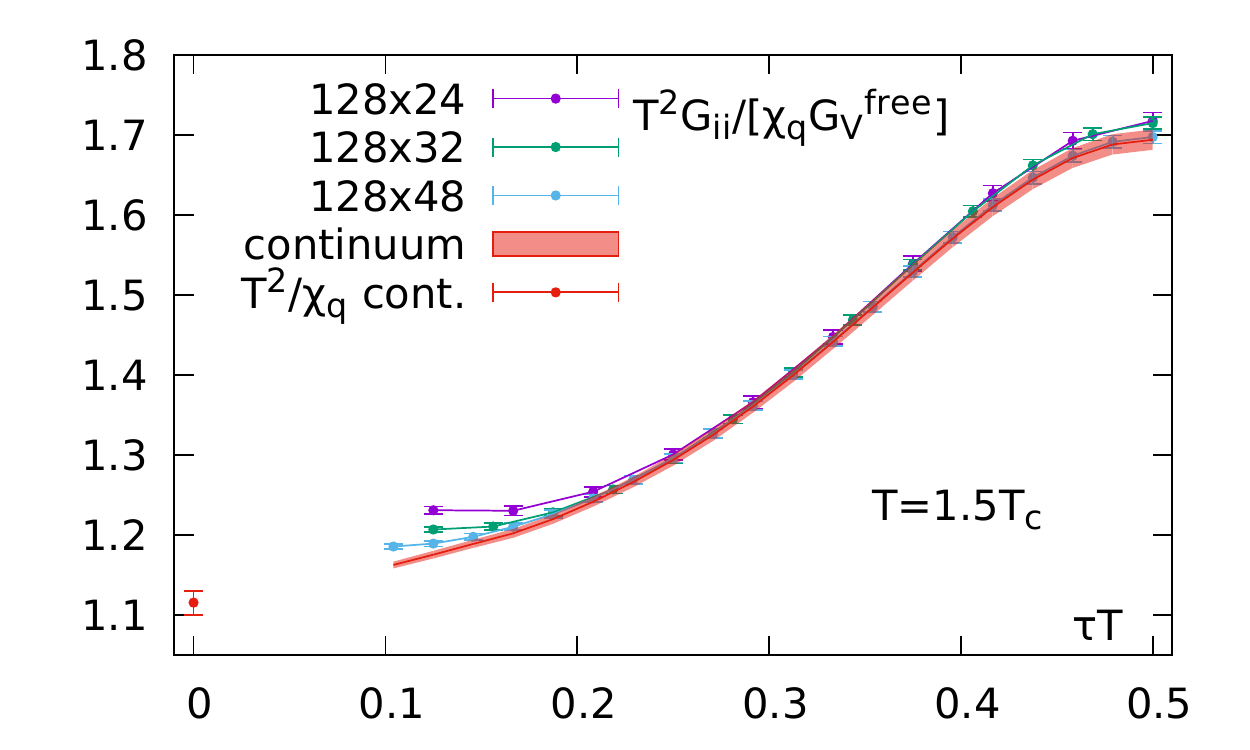}
	\includegraphics[width=0.495\textwidth]{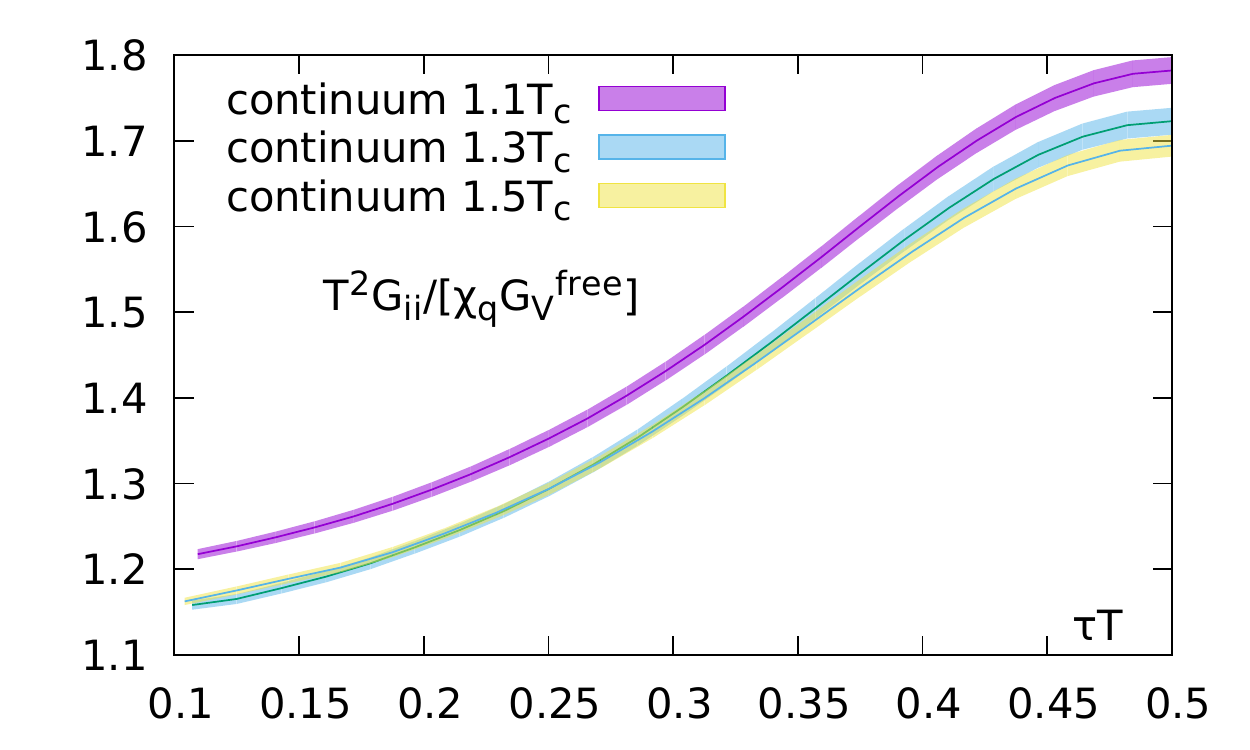}
	\caption{\textit{Left and top right}: All three lattice correlators and the resulting 
		continuum extrapolated correlator for the datasets at $T=1.1T_c$, 
		$T=1.3T_c$ and $T=1.5T_c$, respectively.
		Note that the finest lattice agrees with the continuum extrapolation 
		down to $\tau T\sim0.2$ in all cases. The single black datapoint at
		$\tau T=0$ indicates the continuum extrapolated result for the
		inverted quark number susceptibility.
		\textit{Bottom right}: The continuum extrapolations for all 
	three temperatures.}
	\label{fig_cont_extr}
\end{figure*}
\section{Lattice setup and continuum extrapolation} 
The thermal expectation value of the renormalized Euclidean iso-vector
correlation function,
\begin{eqnarray}
	G_{H}(\tau,\vec{x})&=&\avg{J_{h}(\tau,
          \vec{x})J^{\dagger}_{h}(0,\vec{0})} ,
	\label{eqn_correlator_xspace}
\end{eqnarray}
is constructed from the
renormalized vector current
\begin{eqnarray}
	J_{h}&=&Z_V \bar{\psi}(x)\gamma_{h}\psi(x) ,
\end{eqnarray}
where $Z_V$ is the appropriate renormalization constant, non-perturbatively 
determined in \cite{Luscher1996II},
and $H=hh=00,ii,\mu\mu$ are components of the vector correlation function.
Note that when computing the correlator on the lattice, we do not include 
the sum of squared charges $C_{em}=\sum_fQ_f^2$, and thus the spectral function
obtained from lattice data does not contain this factor, either.\\
The point to point correlators (\ref{eqn_correlator_xspace}) are projected to 
definite momentum $\vec{p}$ by summing over all spatial coordinates,
\begin{eqnarray}
	\label{eqn_correlator_pspace}
	G_H(\tau,\vec{p}) =\sum_{\vec{x}} G_H(\tau,\vec{x})e^{i\vec{p}\vec{x}}.
\end{eqnarray}
In this study we constrain ourselves to the case $\vec{p}=0$. 
Results for non-vanishing momenta and how these allow to obtain lattice constraints
on the thermal photon rate can be found in \cite{Ghiglieri:2016tvj}.\\
Splitting the 
correlation function (\ref{eqn_correlator_pspace}) into spatially and 
temporally polarized components, defining for $H=V$ in Euclidean metric $G_V =G_{ii} + G_{00}$,
we form a ratio of correlation functions 
\begin{eqnarray}
	\label{eqn_ratio}
	R_{ii}&=&\frac{T^2}{\chi_q}\frac{G_{ii}(\tau T)} {G_{V}^{free,lat}(\tau T)},
\end{eqnarray}
\begin{figure}[t]
	\centering
	\includegraphics[width=0.495\textwidth]{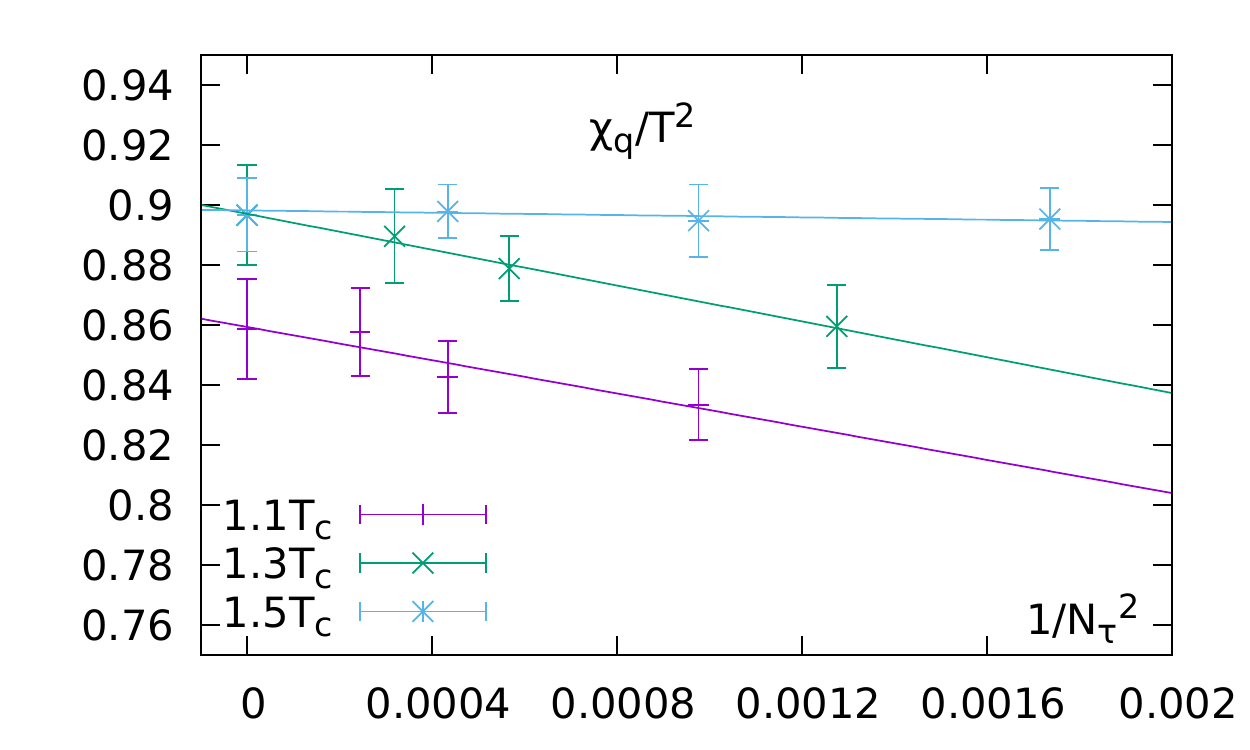}
	\includegraphics[width=0.495\textwidth]{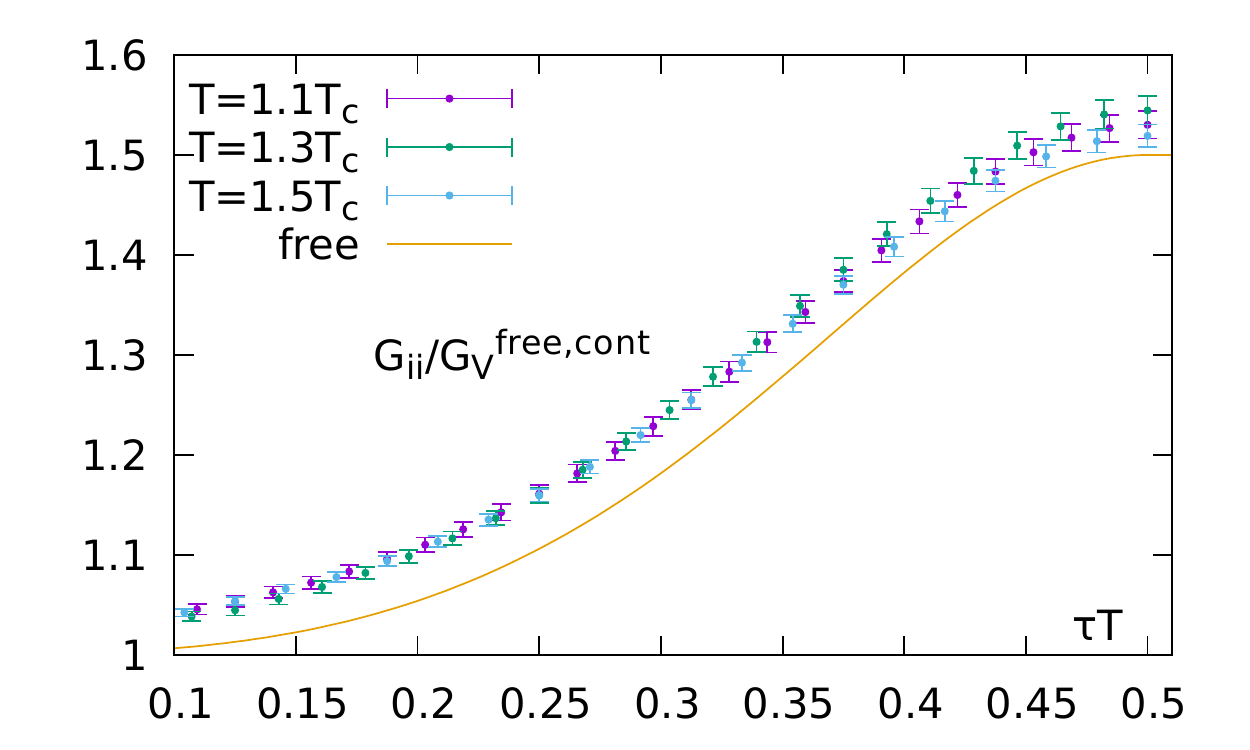}
	\caption{\textit{Top}: the extrapolation of $\chi_q/T^2$. The leftmost datapoints 
	are the continuum extrapolated values.
	\textit{Bottom}: Continuum extrapolated ratios of renormalized correlation 
		functions $G_{ii}/G_V^{\text{free}}$ for
		all three temperatures, not normalized by $\chi_q/T^2$. 
		The solid line is the corresponding non interacting ratio 
	$G_{ii}^{\text{free}}/G_V^{\text{free}}$.}
	\label{fig_corr_degen}
\end{figure}
where $G_{ii}$ is normalized by both the free, massless correlator on the 
lattice \cite{Stickan2003,Karsch2003}, and the 
quark number susceptibility $\chi_q/T^2=-G_{00}/T^3$ 
is defined by the time-time component of the correlation function 
which is constant in Euclidean time.
The division by the latter is used make the the ratio independent 
of renormalization, while
the division by the former cancels the exponential falloff of the interacting correlator.\\
Lattice calculations have been performed in the quenched approximation 
using the standard Wilson gauge action with configurations separated by $500$ 
sweeps of over-relaxed heat bath updates, and 
the non-perturbatively improved Wilson-Clover action \cite{Luscher1996I} for the 
valence quarks.
The bare gauge couplings $g^2(a)=6/\beta$ are fixed to yield 
three different temperatures 
$T=1/(aN_{\tau})=1.1T_c,\;1.3T_c,\;1.5T_c$, with lattice spacings $a$
and lattice time extents
$N_{\tau}$, following the procedure of scale setting recently performed in 
Ref.~\cite{Francis2015}.
The temperatures and scales quoted in Tab.~\ref{tab_lattice_data}
are estimated from the $t_0$-scale \cite{Luscher:2010iy} and $r_0$-scale \cite{Sommer:1993ce} 
determined in \cite{Francis2015}, where also the  
conversions to $T_c$ for the two different scales are based on.
For the 
continuum extrapolated data, in the following, we will state temperatures 
of $T=1.1T_c$, $T=1.3T_c$ and $T=1.5T_c$ for the three data sets, respectively.
For each temperature three increasingly finer lattices are considered to allow for 
linear extrapolations to the continuum limit, see Tab.~\ref{tab_lattice_data}.
Valence quark masses are estimated via the improved Axial Ward Identity (AWI) mass
\cite{Luscher1996I}.
The hopping parameters $\kappa$ are tuned such that the valence quark masses are 
small, corresponding to $m_{\overline{MS}}(\mu=2GeV)\sim \mathcal{O}(10MeV)$ in 
the $\overline{MS}$ scheme.
Note that for the two lowest temperatures the aspect ratio is fixed to
$N_{\sigma}/N_{\tau}=3$ and $N_{\sigma}/N_{\tau}=3.42$ for all lattices, respectively, ensuring a 
constant physical volume, while for the $T=1.5T_c$ lattices the aspect ratio decreases
with decreasing cutoff $a$. However, finite volume effects are supposed to 
be small \cite{Ding2010}.\\
For each of the three temperatures, all three increasingly finer lattices are used to 
perform continuum extrapolations of the ratio (\ref{eqn_ratio}) and the quark number
susceptibility $\chi_q/T^2$. 
In our case, these are linear extrapolations in $1/N_{\tau}^2\sim a^2\longrightarrow 0$, 
as opposed to $1/N_{\tau}\sim a$ for an unimproved Wilson action. 
In order to carry over a maximum of information from the lattices to the continuum,
we interpolate $R_{ii}$ in $\tau T$ on the two coarser lattices using 
a natural cubic spline, 
so that at any distance $\tau T$ available on the finest lattice the continuum limit
can be performed.
The results for the ratio (\ref{eqn_ratio}) are shown in Fig.~\ref{fig_cont_extr}.
The errors on the continuum extrapolated ratios are obtained from a bootstrap analysis,
and are slightly below the one percent level. 
For the renormalized quark number susceptibility $\chi_q/T^2$ we show the extrapolation 
in Fig.~\ref{fig_corr_degen} (\textit{top}) and the resulting continuum extrapolated 
values in Tab.~\ref{tab_contextr_chiq}.
The inverse of the continuum extrapolated susceptibility is 
also shown in Fig.~\ref{fig_cont_extr} (\textit{left}) and (\textit{top right}) as 
red datapoints in the bottom left corners.\\
A comparison of the continuum extrapolated ratios for all three temperatures is shown 
in Fig.~\ref{fig_cont_extr} (\textit{bottom right}). The results for the two highest 
temperatures overlap to a large extent within errors, while the extrapolation for $T=1.1T_c$ lies 
above the former. On the other hand, looking at the continuum extrapolated 
correlator ratio $G_{ii}/G_V^{\text{free}}$, i.e. \textit{without} the dividing 
by $\chi_q$, for 
all three temperatures in Fig.~\ref{fig_corr_degen} (\textit{bottom}), they show a very 
pronounced overlap.
\begin{table}[t]
	\centering
	\begin{tabular}{|c|c|c|c|}
		\hline
		$T$ & $1.1T_c$ & $1.3T_c$ & $1.5T_c$ \\
		\hline
		$\chi_q/T^2$ & $0.857(16)$ & $0.897(17)$ & $0.897(12)$ \\
		\hline
	\end{tabular}
	\caption{The continuum extrapolated values of the quark number susceptibility $\chi_q/T^2$.}
	\label{tab_contextr_chiq}
\end{table}
In the plot the free continuum correlator $G_{ii}^{\text{free}}/G_V^{\text{free}}$ 
is shown as a solid line and is independent of $T$.
The reason for $R_{ii}(1.1T_c)$ deviating 
from the almost overlapping $R_{ii}(1.3T_c)$ and $R_{ii}(1.5T_c)$ is thus the quark number 
susceptibility $\chi_q/T^2$ differing in the two cases, as can be seen from their inverses 
in Fig~\ref{fig_cont_extr}. However, note that this difference 
is rather small. 
From this and the agreement of the correlators in Fig.~\ref{fig_corr_degen}~(\textit{bottom}), the only
scale in this temperature window is the temperature and no
resonance contributions, e.g. a rho meson,  are expected in the gluon plasma in this temperature
region.
Additionally it can already be inferred here that the underlying spectral functions should
be very similar for all three temperatures, already indicating that temperature
effects in the temperature scaled dilepton rates and the electrical
conductivities will be rather small.

\section{Spectral functions \& thermal moments}
\label{sec_fitting}
In order to extract the vector spectral function via 
(\ref{eqn_integ_trans}) we employ an Ansatz for its spatial component,
already used before in \cite{Ding2010},
\begin{eqnarray}
		\label{eqn_ansatz}
		\rho_{\text{ans}}(\omega,T)&=&\chi_q c_{\text{BW}} \frac{\omega\Gamma}{\omega^2+
	(\Gamma/2)^2} \nonumber\\ 
	&+&\frac{3}{2\pi}(1+k)\omega^2\tanh\left(\frac{\omega} {4T}\right)\nonumber\\
	&\equiv& \rho_{\text{BW}}(\omega,T) + (1+k)\rho_V^{\text{free}}(\omega,T).  
\end{eqnarray}
It consists of two constituents: a Breit-Wigner peak, governing the behavior
in the low $\omega$
region, and a modified version of the free, massless continuum spectral function. The 
modification parameter in the latter case fulfills $k=\alpha_s/\pi$ at 
leading order perturbation theory. 
As has been discussed at end of the previous section there is no indication of
resonances at the temperatures studied in this work and therefore we do not include any
bound state contribution in our Ansatz.
The 
Ansatz (\ref{eqn_ansatz}) is inspired by the known relations for massless continuum spectral functions in the 
non-interacting case \cite{Aarts2005}, 
\begin{eqnarray}
	\label{eqn_spf_ii_free}
	\rho_{ii}^{\text{free}}(\omega,T)&=&2\pi T^2\omega \delta(\omega) + \frac{3}{2\pi}
	\omega^2\tanh(\frac{\omega}{4T}), \\
	\label{eqn_spf_00_free}
	\rho_{00}^{\text{free}}(\omega,T) &=& 2\pi T^2 \omega \delta(\omega), \\
	\label{eqn_spf_V_free}
	\rho_{V}^{\text{free}}(\omega,T) &=& \rho_{ii}^{\text{free}}(\omega,T)
	- \rho_{00}^{\text{free}}(\omega,T).
\end{eqnarray}
While the temporal correlator $G_{00}$ also in the interacting case is a constant 
due to charge conservation,
which protects the $\delta$-function contained in its spectral function by symmetry,
the corresponding $\delta$-function in the spatial part is expected
to be washed out upon the onset of interactions
\cite{Hong2010,Moore2006,Aarts2002,Petreczky:2005nh}
reflecting the transport properties of the thermal medium
\cite{Boon_Yip,Forster}.
Following the analysis in \cite{Ding2010}, motivated by arguments from 
hydrodynamics and kinetic theory, 
the $\delta$-function is smeared to a Breit-Wigner peak $\rho_{BW}$
due to thermal modifications. 
{The inverse width of the peak is a measure of the 
correlation time scale of the medium \cite{Meyer2011}. In a quasiparticle 
picture, a larger correlation time
implies that there are fewer (or less strong) interactions to wash out existing
correlations, and hence the medium is more weakly coupled. Concerning the shape of 
the resulting spectral function at low frequencies, 
this is in turn signaled by a narrower peak.\\
An estimator for the spectral function is then obtained from relation 
(\ref{eqn_integ_trans})
by $\chi^2$-minimizing the Ansatz $\rho_{\text{ans}}$ from (\ref{eqn_ansatz})
with respect to the continuum extrapolated ratio data from eqn. (\ref{eqn_ratio}), i.e. 
	\begin{eqnarray}
		\label{eqn_ratio_ansatz}
			R_{ii}(\tau T) &=& \frac{T^2}{\chi_q G_V^{\text{free}}(\tau T)}\int\limits_0^{\infty}
			\!\frac{\dd \omega}{2\pi}\rho_{ii}(\omega,T)K(\omega, \tau, T) \nonumber\\
			&=& \frac{T^3}{2\pi G_V^{\text{free}}(\tau T)}\nonumber\\
                        &&\times\int\limits_0^{\infty}
			\!\text{d}\!\br{\frac{\omega}{T}}\frac{c_{BW}T}{\Gamma}\frac{\omega/T}{\br{\frac{\omega/T}
			{\Gamma/T}}^2+\frac{1}{4}} K(\omega/T, \tau T)\nonumber\\
                    &&+ \frac{T^2}{\chi_q}\br{1+k}.
	\end{eqnarray}
Note that, as we have continuum extrapolated data, the free, massless continuum 
correlation function $G_V^{\text{free}}(\tau T)$, given by
\begin{figure}[t]
	\centering
	\includegraphics[width=0.495\textwidth]{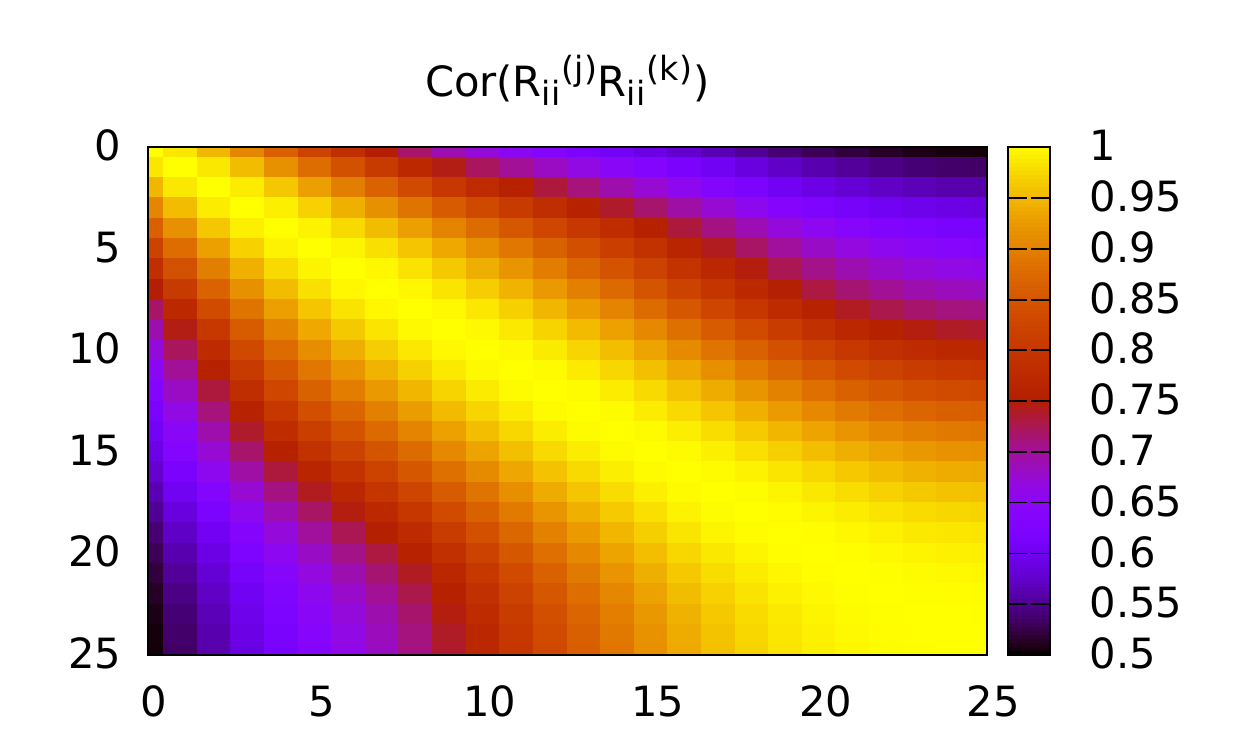}
	\includegraphics[width=0.495\textwidth]{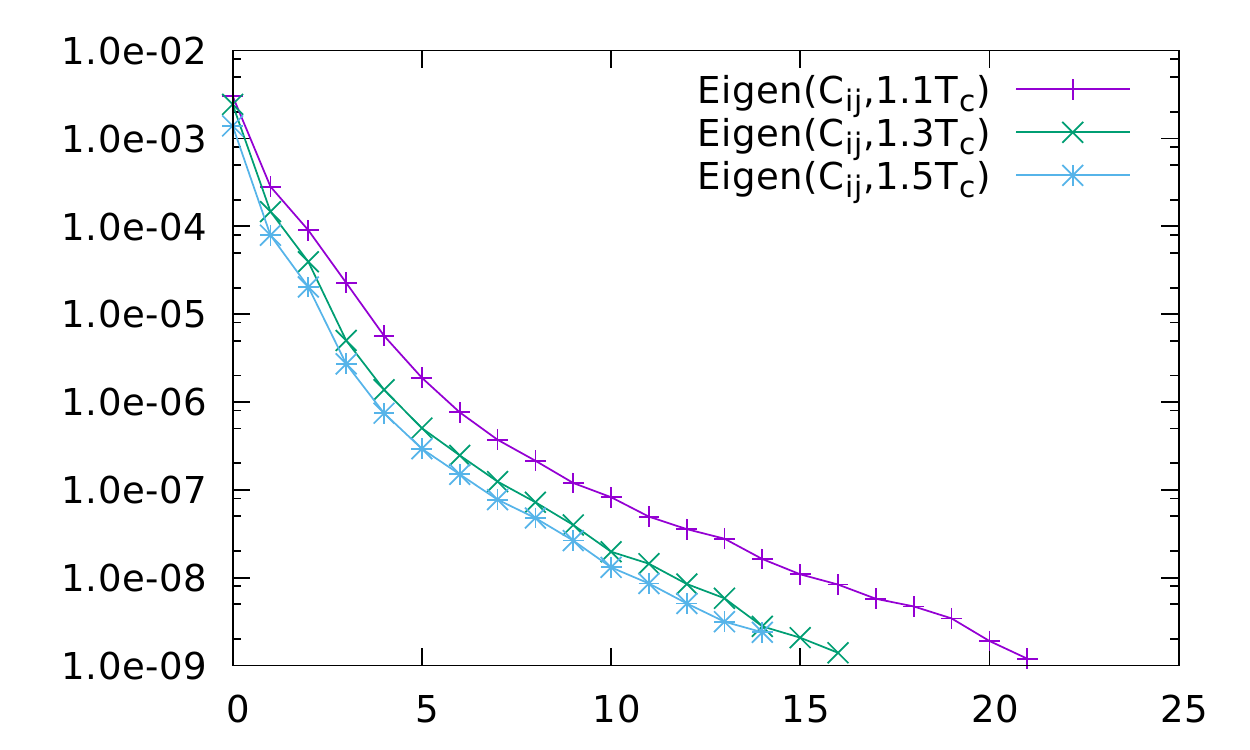}
	\caption{\textit{Top}: A heat map of the entries of the estimated continuum correlation 
		matrix for all points $\tau T > 0.1$ at $1.1T_c$.
		The axes label the row and column entry, resp. Hence, the midpoint 
		$\tau T_j=\tau T_k=0.5$ is located in the bottom right corner.
		\textit{Bottom}: The eigenvalues of the covariance matrices of the data. Note that 
		they decrease in a regular fashion, without strong fluctuations.}
	\label{fig_covar}
\end{figure}
\begin{eqnarray}
		\frac{1}{T^3}G_V^{\text{free}}(\tau T)&=& \frac{1}{2\pi}\int\limits_0^{\infty}
		\!\text{d}\!\br{\frac{\omega}{T}}\rho_V^{\text{free}}(\omega/T) K(\omega,\tau T)\nonumber\\
		&=&6\bc{\pi\br{1-2\tau T}\frac{1+\cos^2(2\pi\tau T)}{\sin^3(2\pi\tau T)}\right.\nonumber\\
   &&	\left.  +2\frac{\cos(2\pi\tau T)}{\sin^2(2\pi\tau T)}},
\end{eqnarray}
appears in the r.h.s. after the integration over $\omega/T$ is performed.
It thus cancels with the normalizing free spectral function and 
the free part of the Ansatz simplifies to a constant in the fit.
The fit itself is performed by taking into account all statistical correlations among the 
data points,
with the covariance matrix $\mat{M}$ of the extrapolated continuum ratio $\hat{R}_{ii}(\tau T)$ 
being estimated from $N_{bs}$ available bootstrap samples 
$R^{(n)}_{ii}(\tau T)$ via the bootstrap 
estimator
\begin{eqnarray}
		M_{jk} = \frac{1}{N_{bs}}\sum_{n=1}^{N_{bs}} &\br{R_{ii}^{(n)}(\tau T_j)-\hat{R}_{ii}(\tau T_j)}\nonumber\\
		\times &\br{R_{ii}^{(n)}(\tau T_k)-\hat{R}_{ii}(\tau T_k)}.
\end{eqnarray}
From the entries of the correlation matrix 
\begin{eqnarray}
	C_{ij}=\frac{M_{ij}}{\sqrt{M_{ii}M_{jj}}}, \qquad \text{(no sum)}
\end{eqnarray}
visualized in Fig.~\ref{fig_covar} (\textit{top}) for the case $T=1.1T_c$, 
it becomes apparent that 
there are statistical correlations between data points throughout the 
whole range of $\tau T > 0.1$. 
Although they fall off with rising distance, they are still larger than $0.5$ 
for all points that will be considered in the fit
and hence non-negligible in the construction of the $\chi^2$ function. This is in accordance
with \cite{Ding2010}, where the $1.5T_c$ dataset has been used for a similar procedure. There the 
fit is done with only the diagonal parts of the covariance matrix and it yields 
a very small value of $\chi^2/\text{dof}$, which is due to neglecting 
correlations among the data. Fig.~\ref{fig_covar} (\textit{bottom}) shows the eigenvalues of 
the used covariance matrices. Note that the estimated condition number 
$\sigma_{\text{max}}/\sigma_{\text{min}}\sim10^6$, with $\sigma_{\text{min/max}}$ being 
the smallest/largest eigenvalue of the covariance matrix,
is large, but sufficiently small for a stable inversion of $C_{ij}$.\\
However, the information about the small $\omega$ 
region resides in the large $\tau T$ region of the correlator 
\cite{Aarts2002II}, i.e.
around its midpoint. In order to extract more information from this region
we also extract the second thermal moment of the correlator data and account for it in 
the fit procedure as an additional data point. The thermal moments are
defined by Taylor expanding around the midpoint,
\begin{eqnarray}
	\label{eqn_thermal_moment_expansion}
	G_H(\tau T) &=& \sum_{n=0}^{\infty}G^{(n)}_H\left(\frac{1}{2}-\tau,
          T\right)^n 
\end{eqnarray}
where
\begin{eqnarray}
		G_H^{(n)}&=&\left.\frac{1}{n!}\frac{d^nG_H(\tau T)}{d(\tau T)^n}\right|_{\tau T = 1/2} \\
		 &=& \frac{1}{n!}\int\limits_0^{\infty}\frac{\dd \omega}{2\pi}\left(\frac{\omega}{T}\right)^n
		\frac{\rho_H(\omega)}{\sinh(\omega/(2T))}.
	\label{eqn_thermal_moment}
\end{eqnarray}
\begin{figure}[t]
	\centering
	\includegraphics[width=0.495\textwidth]{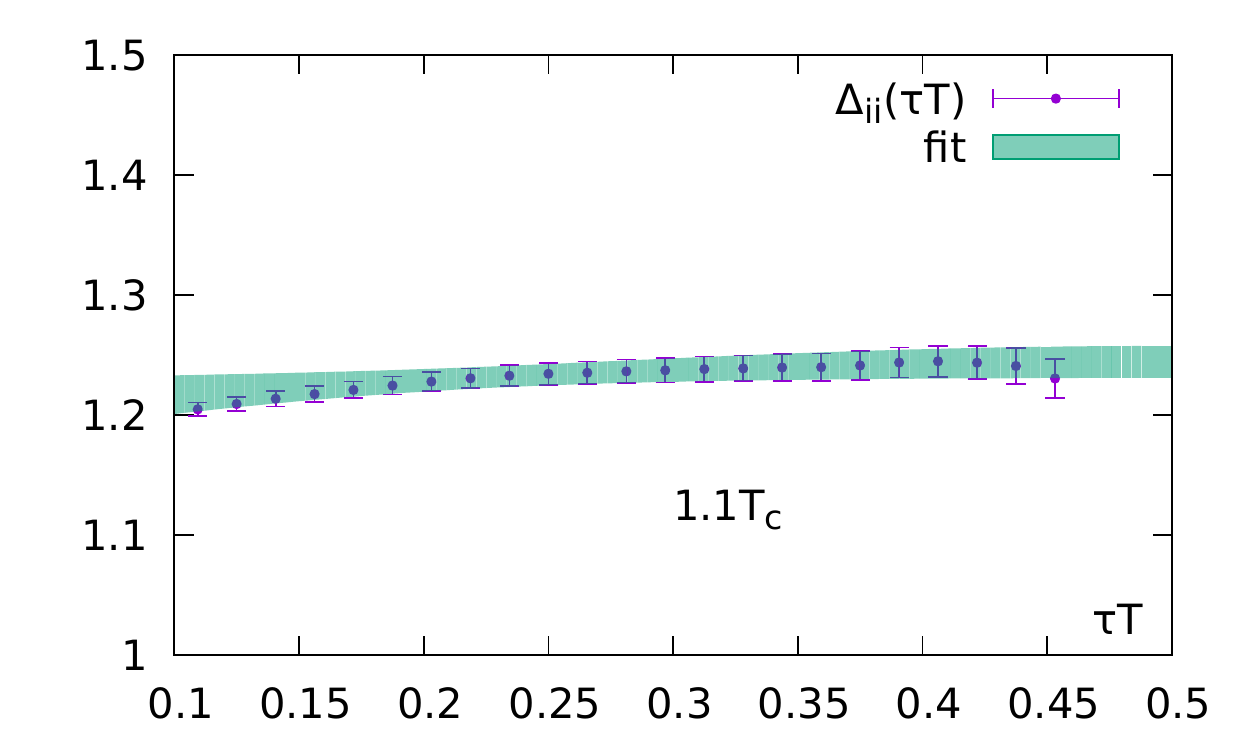}
	\caption{The necessary extrapolation in $\tau T$ to obtain $\Delta_{ii}$
		for the case $T=1.1T_c$.
		The fit interval is $\tau T\in[0.2,0.45]$, i.e. the point at the far 
	right is not included in the fit.}
	\label{fig_delta_extr}
\end{figure}
Due to the symmetry of the integral, the odd thermal moments 
$G_{H}^{(2n+1)}$ vanish. 
The first thermal moment $G_H^{(0)}$ is the value of the correlation function at the 
midpoint, which is included in the fit trivially, while the second thermal moment 
$G_{H}^{(2)}$
is the curvature of the correlation function at the midpoint.  
In order to extract it from the lattice data, we further define the midpoint 
subtracted correlator ratio by
\begin{eqnarray} 
	\label{eqn_midp_subtract}
	\Delta_H(\tau T)\equiv \frac{T^2}{\chi_q}\frac{G_H(\tau T) - 
	G_H^{(0)}}{G^{free}_H(\tau T) - G_H^{(0),free}},
\end{eqnarray}
which at the midpoint satisfies
\begin{eqnarray} 
	\label{eqn_midp_subtract_eval_midp}
	\Delta_H(\tau T)\overset{\tau T \rightarrow 1/2}{\longrightarrow}
	\frac{T^2}{\chi_q}\frac{G_H^{(2)}}{G_H^{(2),free}}\;, 
\end{eqnarray}
hence arriving at a ratio similar to (\ref{eqn_ratio}).
Because we cannot compute the limit (\ref{eqn_midp_subtract_eval_midp}) 
directly from the lattice data, we first compute (\ref{eqn_midp_subtract}) 
for each lattice spacing for all available distances $\tau T<0.5$, 
and extrapolate this to the continuum limit. From (\ref{eqn_midp_subtract}) and 
the expansion (\ref{eqn_thermal_moment_expansion}) we find an Ansatz to extrapolate
the resulting continuum extrapolated midpoint subtracted correlator ratio 
to $\tau T=0.5$,
	\begin{eqnarray}
		\label{eqn_midp_subtr_expansion}
		\Delta_{ii} = \frac{T^2}{\chi_q}\frac{G_{ii}^{(2)}}{G_{ii}^{(2),free}} 
		\Bigg\{1 &+&
                  \br{R^{(4,2)}_{ii}-R^{(4,2)}_{ii,free}}\br{\frac{1}{2}-\tau
                    T}^2 \nonumber\\
	 &+& \ \mathcal{O}\br{\br{\frac{1}{2}-\tau T}^4}\Bigg\}, \\
	\nonumber \text{with} \ \ R_{X}^{(n,m)} &=& G^{(n)}_{X} / G^{(m)}_{X}. 
\end{eqnarray}
The two unknown parameters are the thermal moment 
$G_{ii}^{(2)}/G_{ii,free}^{(2)}$
and the ratio $R_{ii}^{(4,2)}$. Fig.~\ref{fig_delta_extr} shows the extrapolation for the 
case $T=1.1T_c$. The results for $\Delta_{ii}$ and the ratio $R_{ii}^{(4,2)}$ are 
shown in Tab.~\ref{tab_delta_extr}.
Note that since the obtained value for the ratio $R_{ii}^{(4,2)}$ 
is quite accurate, we could also use it to constrain the fits.
However, looking at (\ref{eqn_thermal_moment}), we see that the maximum
\begin{figure*}[t]
	\centering
	\includegraphics[width=0.495\textwidth]{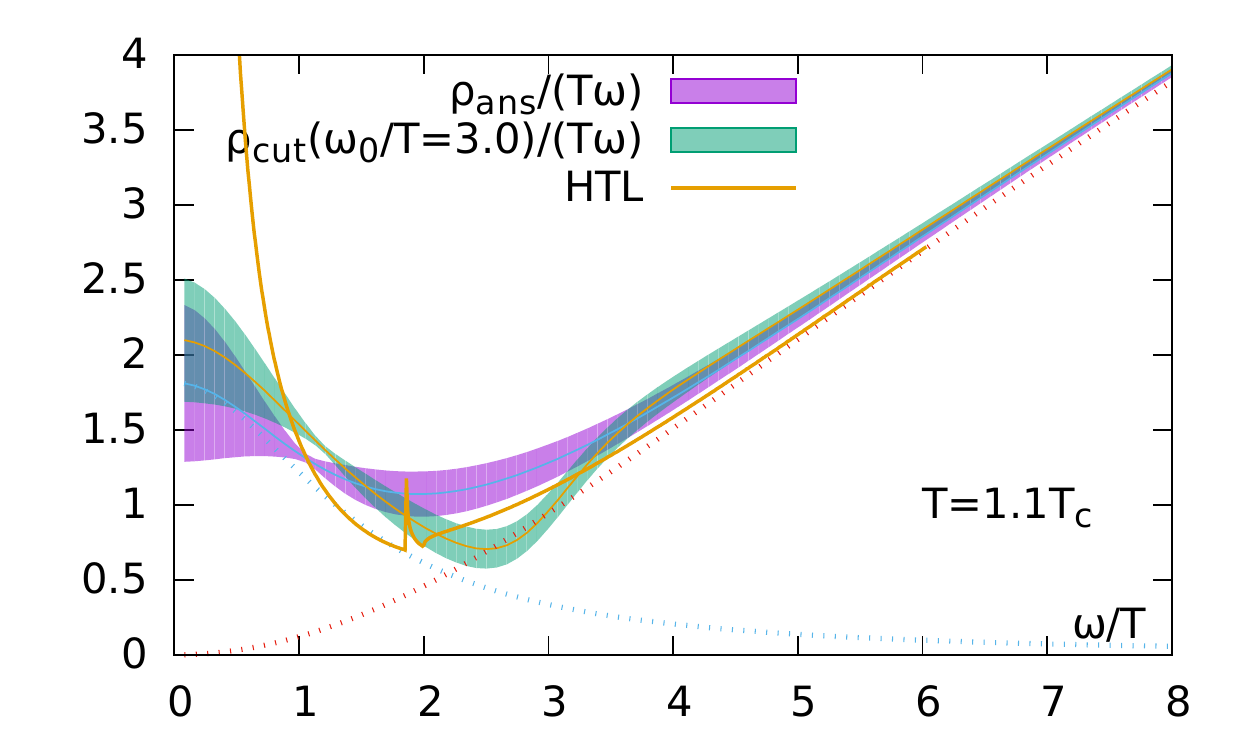}
	\includegraphics[width=0.495\textwidth]{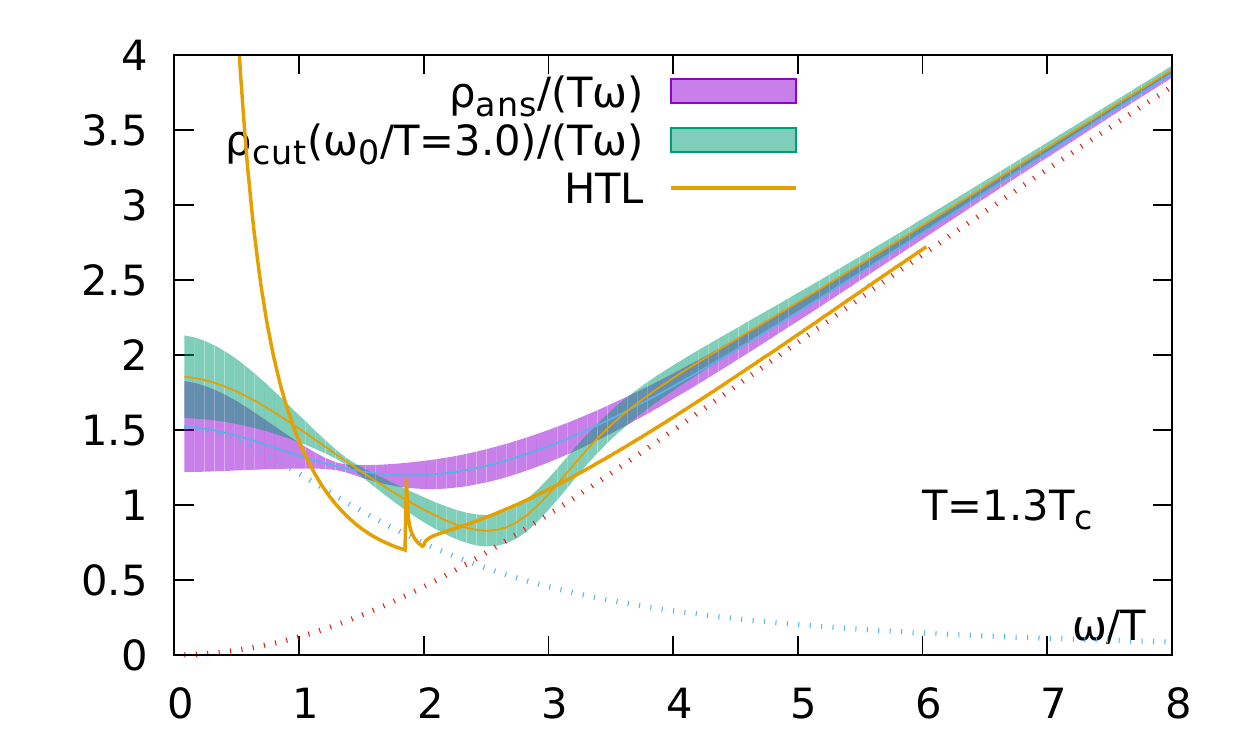}
	\includegraphics[width=0.495\textwidth]{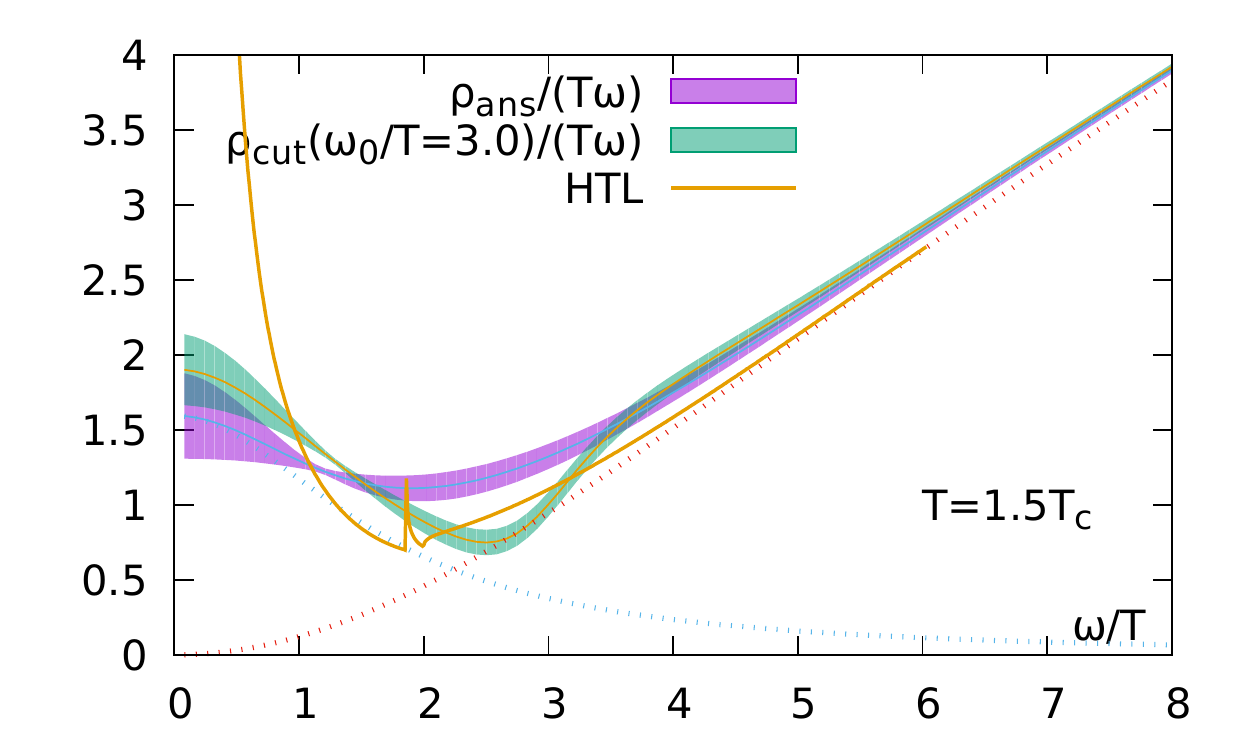}
	\includegraphics[width=0.495\textwidth]{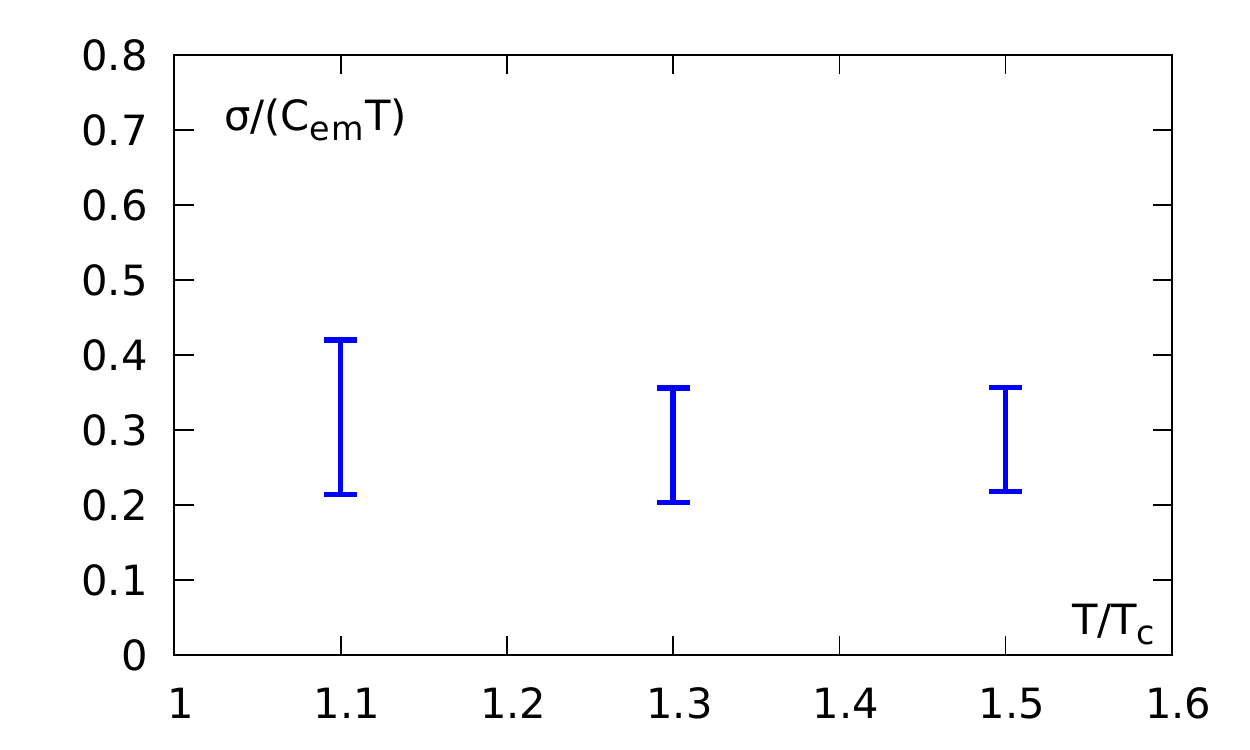}
	\caption{The spectral functions resulting from the fit for
		all temperatures. The dotted lines are the Breit-Wigner and the free contributions
		separately to guide the eye. Note the consistently higher intercept 
		of the spectral functions with the cut applied.
		\textit{Bottom right:} The final results for the electrical conductivity for 
	all three temperatures as resulting from Ansatz $\rho_{\text{ans}}$.}
	\label{fig_spf_plot}
\end{figure*}
of the weight $\frac{\br{\omega/T}^n}{n!\sinh(\omega/(2T))}$ in the integrand
is given by the self-consistent expression 
$\omega/T=2n\tanh(\omega/(2T))$, which means that for larger $n$ the 
region of dominating weight shifts to larger frequencies. In turn, this 
moves the focus further away from the region of interest.\\
As a remark, it would in principle be 
easier to obtain the second moment from the continuum extrapolated 
ratio data (\ref{eqn_ratio})
in a similar fashion by expanding $G_{ii}(\tau T)/G_{ii}^{\text{free}}$ in
terms of moments, as we already have it at zero lattice spacing and no additional 
continuum extrapolation would be necessary.
\begin{table}
	\begin{tabular}{|c||c|c|c|}
		\hline 
		$T$ & $\Delta_{ii}$ & $R^{(4,2)}_{ii}$ & $\chi^2/\text{dof}$ \\
		\hline
		$1.1T_c$ & $1.245(14)$ & $10.287(120)$ & $0.67$ \\
		$1.3T_c$ & $1.203(12)$ & $10.070(108)$ & $0.50$ \\
		$1.5T_c$ & $1.200(9)$ & $10.205(73)$ & $0.89$ \\
		\hline
	\end{tabular}
	\caption{The results for the extrapolation in $\tau T$ to obtain 
		$\Delta_{ii}$. The fits have been performed using the full covariance 
	matrix of the data.}
	\label{tab_delta_extr}
\end{table}
The advantage in constructing the midpoint subtracted correlator ratio from the data 
from scratch is that the desired second thermal moment 
then appears in the Ansatz (\ref{eqn_midp_subtr_expansion}) as its intercept, while expanding 
analogously for $G_{ii}(\tau T)/G_{ii}^{\text{free}}$ would 
yield the second thermal moment as its curvature. The former is simply more reliable to 
obtain from a fit.\\

\section{Analyses of vector current correlation functions}
When estimating the systematics of our procedure, 
an essential source of uncertainty is the fit Ansatz itself.
Because of the general lack of information in an ill-posed
inversion problem, and the fact that we add information 
by choosing our Ansatz, which is inspired from phenomenology,
it is not excluded that other Ans\"atze
fit the data as well. In the next section we thus complement
the analysis of Ansatz $\rho_{\text{ans}}$ by developing several structural changes,
and discuss what conclusion could be drawn from the respective modified 
Ansatz. Finally, the fit procedure is applied using each new Ansatz, and 
the results are presented.\\

\subsection{Spectral function Ansatz: Breit-Wigner peak + free continuum}
In the fit of our Ansatz $\rho_{\text{ans}}$ to the extrapolated continuum data we 
want to provide as much physical information as possible.
From the continuum extrapolations shown in 
Fig.~\ref{fig_cont_extr} we see that for all three temperatures 
the extrapolation results almost 
agree with the data on the corresponding finest lattice from the midpoint 
down to $\tau T \simeq 0.15-0.20$. This is also where the coarsest 
lattice starts to bend upwards. As the ratios are supposed to approach 
$R_{ii} \longrightarrow T^2/\chi_q$ in the limit $\tau T \longrightarrow 0$, 
the 'bending up' when going to shorter distances is a cutoff
effect. Since we want to be sure to include only
continuum data, which is free of those effects, into our fit procedure, 
we aim for $\tau T\sim0.2$ and in practice take the distance which 
yields the $\chi^2/\text{dof}$ closest to unity when fitting 
Ansatz $\rho_{\text{ans}}$.
This amounts to $\tau T_{\text{min}}=0.187,0.232,0.229$ for $T=1.1T_c,1.3T_c,1.5T_c$,
respectively, which will also be used as a definite choice of fit intervals 
for all following fits.\\
The fits of $\rho_{\text{ans}}$ to the continuum extrapolated correlator data
show a good convergence behavior and 
yield as a result the three fit parameters $\Gamma$, $c_{BW}$, $k$
and their respective statistical fit errors, see Tab.~\ref{tab_fit_results_ans}.
The relative statistical fit errors of the parameters are roughly 
$25-40\%$ for $c_{BW}T/\Gamma$ and $20-30\%$ for $\Gamma/T$. Note that the former 
has been calculated taking into account the correlation of the two 
parameters. The dimensionless modification $k$ to the large frequency free behavior 
is small, but distinctly larger than zero in all cases. 
Their values are reasonable when using the leading order perturbative
relation,
$k\simeq \alpha_s/\pi$, to compare to other determinations
of temperature dependent running couplings \cite{Kaczmarek:2004gv}.
However, within errors there 
is no visible trend for the available range of temperatures.
At this point we treat $k$ or $\alpha_s$ as a constant and do not include a
dependence on $\omega$, i.e. a running $\alpha_s$. In the following subsections we
will study a class of spectral functions and also model the effect of more
involved perturbative large $\omega$ behavior in sec.~IV.D. 
The values of $\chi^2/\text{dof}$ vary around unity and show that the fit to the 
data performs reliably.
Using these parameters and their correlation matrix we construct
the resulting spectral function, normalized by the frequency, with its 
corresponding statistical 
errorband in Fig.~\ref{fig_spf_plot} (\textit{left}) and (\textit{top right}). The $HTL$ result 
\cite{Braaten1990}
is also plotted and lies mostly below our estimate 
of the spectral function. 
In the low frequency region a straightforward HTL-resummation
can not reproduce a finite electrical conductivity as it 
behaves as $\rho_{HTL}\sim 1/\omega$ for small $\omega$
\cite{Braaten1990,Karsch:2000gi}.\\ 
The electrical conductivity is obtained from the origin of the 
spectral function via the Kubo relation (\ref{eqn_kubo}), which is 
proportional to the ratio of the fit parameters $C_{\text{BW}}$ and 
$\Gamma/T$, written as follows,
\begin{eqnarray}
	\frac{\sigma}{C_{em}T} =  \frac{2}{3T}\chi_q\frac{c_{BW}}
	{\Gamma}.
\end{eqnarray}
As an intermediate step in our analysis, it is given in Tab.~\ref{tab_fit_results_ans}
for all three temperatures with the corresponding fitting error.
Consequently, the soft photon rate can also be obtained and written in 
terms of the electrical conductivity as follows,
\begin{eqnarray}
	\label{eqn_photonrate_conductivity}
	\lim_{\omega\to0}\omega\frac{\dd R_{\gamma}}{\dd^3p}
	=\frac{\alpha_{em} C_{em}}{2\pi^2}\br{\frac{\sigma}{C_{em}T}}T^2,
\end{eqnarray}
and is presented at the end of this work, including the systematics developed 
in the following sections.\\
\begin{table}[t]
	\begin{tabular}{|c||c|c|c|c|c|}
		\hline
		$T$ & $\sigma/(C_{\text{em}}T)$ & $\Gamma/T$ & $c_{BW}T/\Gamma$ & $k$ & $\chi^2/\text{dof}$ \\
		\hline
		$1.1T_c$ & $0.302(88)$ & $2.86(1.16)$ & $0.528(154)$ & $0.038(8)$ & $1.15$ \\
		$1.3T_c$ & $0.254(51)$ & $3.91(1.25)$ & $0.425(85)$ & $0.029(9)$ & $0.52$ \\ 
		$1.5T_c$ & $0.266(48)$ & $3.33(89)$ & $0.445(80)$ & $0.040(7)$ & $1.13$ \\
		\hline
	\end{tabular}
	\caption{Results of fitting the Ansatz $\rho_{\text{ans}}$ for all three temperatures.}
	\label{tab_fit_results_ans}
\end{table}

\subsection{Spectral function Ansatz: flat transport peak + free continuum}
The Ansatz above is motivated by kinetic theory computations and arguments.
On the other hand, in the strong coupling limit the vector spectral 
function can be obtained from the AdS/CFT correspondence. The resulting 
spectral function in the low frequency region usually has no peak 
structure \cite{Teaney2006}, consisting
of a flat, 'featureless' shape in $\rho/\omega$ and then going over
into a typical large frequency behavior. This transition
is typically accompanied by small, exponentially damped oscillations.
A simple Ansatz roughly showing this behavior is given by
\begin{figure}[t]
	\centering
	\includegraphics[width=0.495\textwidth]{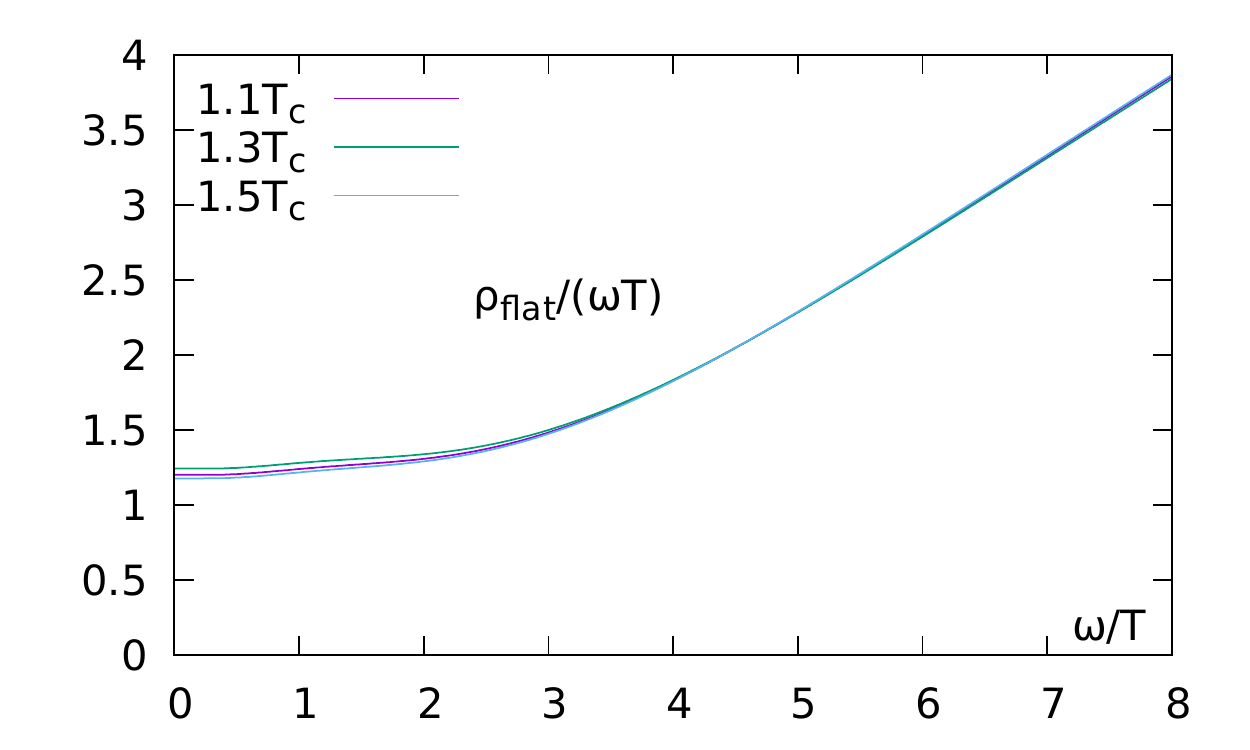}
	\caption{The spectral function resulting from the fit of the (coarse) 
	model $\rho_{\text{flat}}$ for all temperatures.}
	\label{fig_adscft_fit}
\end{figure}
\begin{eqnarray}
	\label{eqn_model_adscft_theory}
		\rho_{\text{flat}}(\omega)=&a\chi_q\omega \br{1-\widetilde{\Theta}(\omega_0,\Delta_0)}\nonumber\\
								   &+(1+k)\rho_{\text{free}}(\omega)\widetilde{\Theta}(\omega_1,\Delta_1),
\end{eqnarray}
with $\omega_i$ and $\Delta_i$ chosen such that $\rho/\omega$ then results 
in the desired shape. 
The functions 
$\widetilde{\Theta}(\omega_i,\Delta_i)$ are smoothed Heaviside functions 
\begin{eqnarray} 
	\label{eqn_cut_mod}
		\widetilde{\Theta}(\omega,\omega_i,\Delta_i) &=& \br{1+\exp\br{\frac{\omega_i^2-\omega^2}{
	\omega\Delta_i}}}^{-1},
\end{eqnarray}
which become sharp Heaviside functions in the limit $\Delta_i\to0$.
The cut on 
the first term is needed to make sure the large frequency regime is not 
affected by the low frequency constant contribution.
This is of course
a very rough model: not only is there a certain arbitrariness in the choice of 
$\omega_i$ and $\Delta_i$, but in general there are many possible expressions
that approximately describe the desired functional shape.
Also, details like the exponentially damped oscillations are not built into this model.
For these reasons we do not give definite results for the electrical conductivity or 
the photon rate, 
and merely utilize the model to 
test a non-peaked, flat low frequency region in $\rho/\omega$. 
Technically, this change of the Ansatz, compared to the previous case, aims
at making a statement about the \textit{resolution} of our fit method regarding the 
low frequency region of the spectral function.\\
\begin{figure*}[t]
	\centering
	\includegraphics[width=0.495\textwidth]{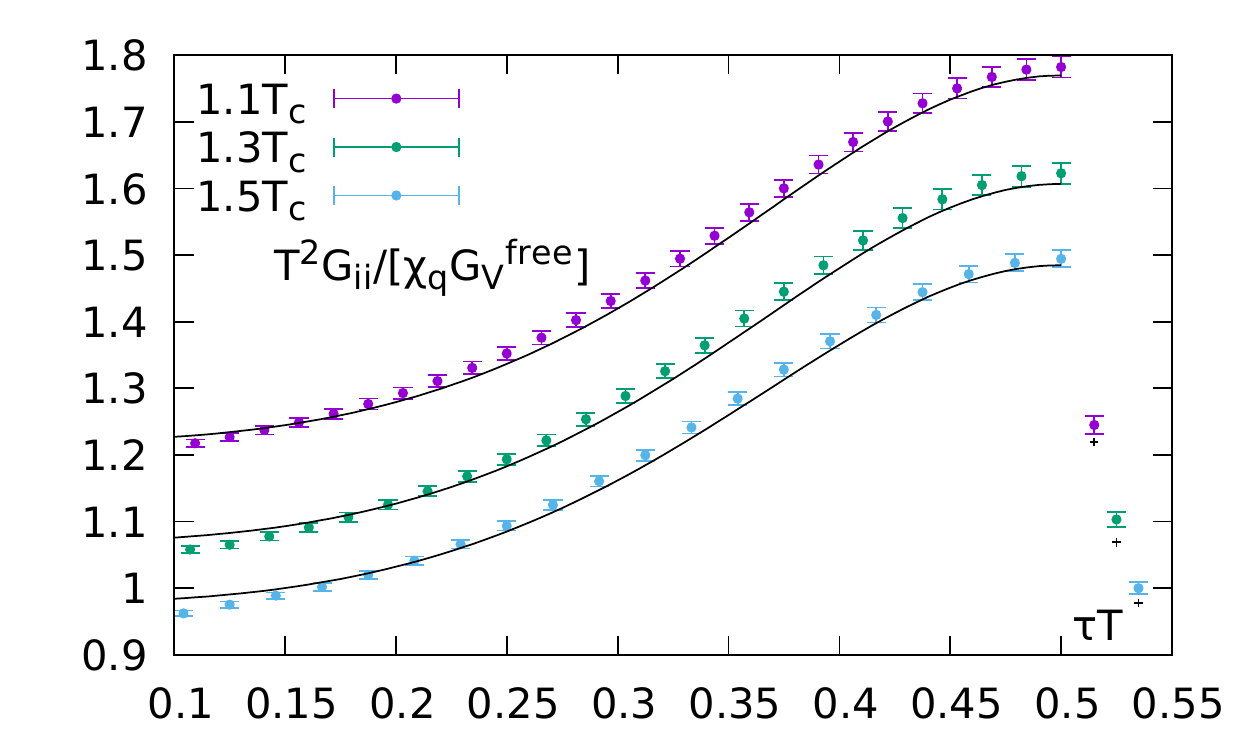}
	\includegraphics[width=0.495\textwidth]{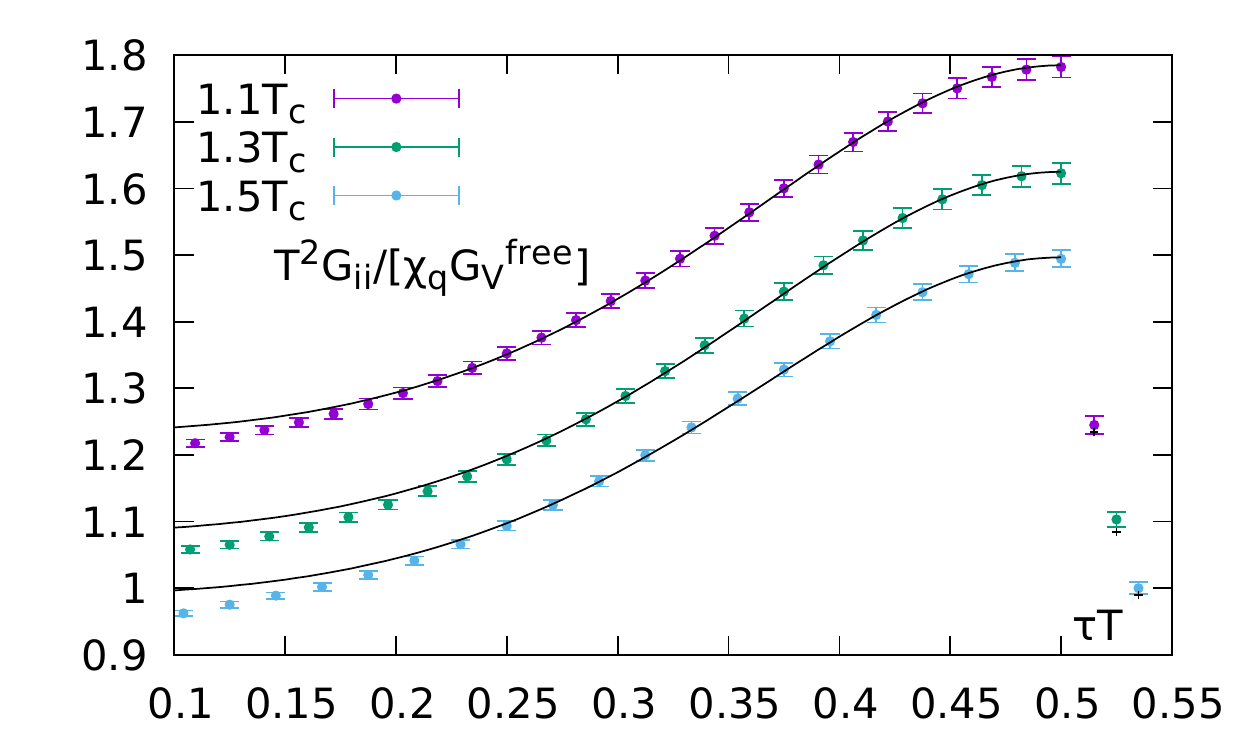}
	\caption{Fit of a real delta peak in the low frequency region. The
		points at $\tau T>0.5$ are the second thermal moments and their fit results, 
		respectively. 
		\textit{Left}: fit including the covariance of the data. Note how the 
		second thermal moments are described much worse than 
	the corresponding correlator data points. 
	\textit{Right}: fit without the covariance of the data. The correlator 
	is described nicely, the thermal moments are a bit off, but less than in 
	the fit including the covariance. The curves are offset for visibility.}
	\label{fig_corr_rec_realdelta}
\end{figure*}
When fitting $\rho_{\text{flat}}$ to the data, we tune the cut 
positions $\omega_i$ and the smoothing parameters 
$\Delta_i$ in such a way that the result from the fit roughly 
describes the characteristic, featureless ADS/CFT solution. The fits work well
for a range of cuts at $\omega_i$ and smoothing parameters $\Delta_i$.
Throughout all temperatures they yield good fit qualities 
of $\chi^2/\text{dof}\sim1.1$ for $T=1.1T_c$,$1.2T_c$ and 
$\chi^2/\text{dof}\sim0.5$ for $T=1.2T_c$, see 
Fig.~\ref{fig_adscft_fit} for the resulting spectral functions.
The interpretation of this is first, that qualitatively this type 
of solution, being featureless in the low frequency region, 
fits our data just as well as a broad Breit-Wigner
peak, motivated by kinetic theory, does.
This implies that our method, with regard to the available data, 
does not have the resolution to differentiate between these two shapes.
The second point to make is 
that when varying the cut positions in such a way that we still have 
a smooth curve, we always end up with an electrical 
conductivity that is close to the lower bound of the results presented in 
Tab.~\ref{tab_fit_results_ans}, i.e. when using $\rho_{\text{ans}}$.\\

\subsection{Crosscheck at low frequency}
\label{sec_squeezing_BW}
As a rather technical crosscheck, instead of using a Breit-Wigner peak for the low 
frequency part of the spectral function, we change it to be a 
real $\delta$ function with variable height, 
\begin{eqnarray}
	\label{eqn_realdelta_mod}
	\rho_{\delta}(\omega)=
	a\chi_q\omega\delta(\omega) + (1+k)\rho_V^{\text{free}}(\omega).
\end{eqnarray}
Up to the parameter $k$ this is essentially the free case. 
Theoretically, when turning off interactions, the conductivity should 
approach infinity, since no force changes the state 
of motion of a charge. Using the Kubo formula, this is clearly reflected in 
the above Ansatz $\rho_{\delta}$ for $\omega\to0$, i.e. it is incompatible with a 
finite conductivity.
Thus, performing the fit using this Ansatz we can check whether this wrong assumption
works out with our interacting data, which should definitely yield a \textit{finite}
conductivity.\\
Performing the fit with Ansatz $\rho_{\delta}$ we find that 
the procedure yields values of $\chi^2/\text{dof}\sim1.5$ for 
the two lower temperatures, and $\chi^2/\text{dof}\sim2.5$, for $1.5T_c$, which 
also quantitatively shows a decrease in fit quality.
Looking at the resulting correlators, shown in 
Fig.~\ref{fig_corr_rec_realdelta} (\textit{left}) for all temperatures, we 
see that the reconstructed curves really underestimate respective the correlator 
data points 
systematically by an amount of one standard deviation or more. Specifically,
the fitted second thermal moments, shown at $\tau T=0.535$ in the plot, drastically 
deviate from the data. 
We conclude that the Ansatz does not describe the data sufficiently, and also place
an emphasis on the importance of accurately
determined thermal moments for the analysis. However, 
one peculiarity in this case is that, when we perform the fit \textit{without}
the covariance matrix in the minimizing $\chi^2$ term, i.e. minimizing only 
with respect to the diagonal terms, we end up with a function
that reconstructs the data points reasonably well at large distances,
see Fig.~\ref{fig_corr_rec_realdelta} (\textit{right}),
and also shows the usual small
$\chi^2/\text{dof}\sim \Obs(0.1)$, which is typical for missing correlations.
In this case, the second thermal moment is not quite as well 
reproduced compared to the data points of the ratio $R_{ii}(\tau T)$, but still 
distinctly better than in the fully correlated case.
Reversing the argument, we see that the a priori insufficient fit Ansatz 
$\rho_{\delta}$, which yields no finite conductivity,
fails to describe the data \textit{only} if the information of the full covariance matrix 
is incorporated in 
the fit. In this sense we find that including covariances in the fit procedure 
measurably enhances our resolution of the spectral function in the low
frequency region.\\

\subsection{Uncertainties from the high frequency region}
\label{sec:uncert}
In order to check for uncertainties arising from the way we model the 
high frequency behavior in $\rho_{\text{ans}}$,
we introduce a low-frequency cutoff multiplied to $\rho_V^{\text{free}}$,
as proposed in \cite{Ding2010},
so that in total the modified Ansatz is given by
\begin{eqnarray}
		\rho_{\text{cut}}(\omega,\omega_0,\Delta_0) &=&\rho_{\text{BW}}(\omega)\nonumber\\
		&+& (1+k)\rho_{V}^{\text{free}}(\omega)\widetilde{\Theta}(\omega,\omega_0,
		\Delta_0).
\end{eqnarray}
The cutoff factor $\widetilde{\Theta}(\omega,\omega_0,\Delta_{\omega})$ is a representation 
of the Heaviside function for
$\Delta_{\omega}\longrightarrow0$, see equation (\ref{eqn_cut_mod}).
Consider that our choice in $\rho_{\text{ans}}$ to account for the large 
frequency regime is essentially the free vector spectral function.
However, this function has positive contributions for all positive frequencies 
$\omega>0$, and it influences the Breit-Wigner peak for small frequencies.
Thus we probe for this influence by cutting off the low frequency part and 
observing how the fit results react on this.\\
\begin{figure}[t]
	\centering
	\includegraphics[width=0.495\textwidth]{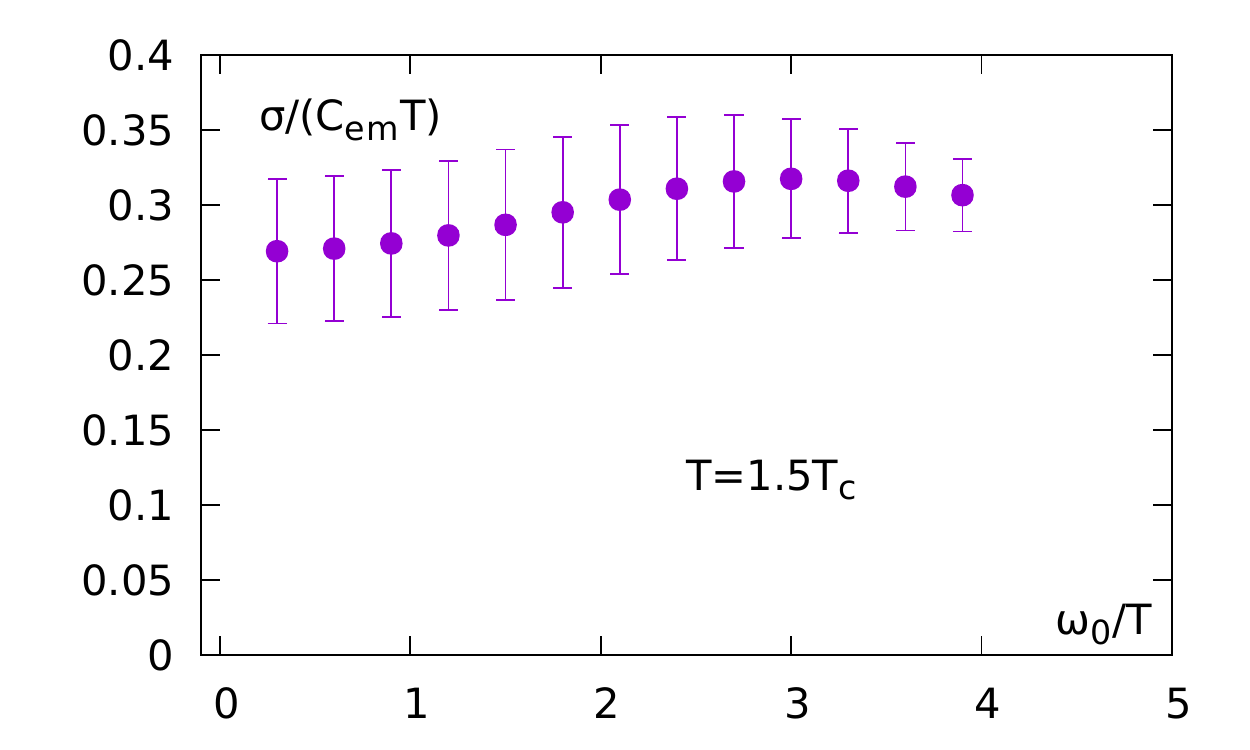}
	\caption{The increase of electrical conductivity upon increase of the 
		cutoff $\omega_0/T$ in (\ref{eqn_cut_mod}). It reaches its maximum
		around $\omega_0/T\simeq3$
		for all three temperatures. The smoothing parameter is fixed to
	$\Delta_0/T=0.5$ throughout the analysis.}
	\label{fig_sigma_rise}
\end{figure}
In order to fit the function $\rho_{\text{cut}}$ to the continuum extrapolated data,
we first of all set the 
width of the smeared Heaviside function to $\Delta_0/T=0.5$. We varied the 
value of $\Delta_0/T$ and found that the result 
does not strongly depend on it. Applying the cut to 
different frequencies $\omega_0/T$ has direct effect on the 
resulting electrical conductivity,
illustrated in Fig.~\ref{fig_sigma_rise}.
As can be seen, it
rises slightly when moving the cut to
higher frequencies, showing that the peak rises in height. Around 
$\omega_0/T \simeq 3$ also $\Gamma/T$ starts to rise sharply, i.e. at that point 
the peak is becoming much broader to compensate for
the missing free contribution in the low $\omega$ regime, and thus
$c_{BW}T/\Gamma$ falls off again. The fit itself still works well over a long range 
of $\omega_0$ in the sense that 
$\chi^2/\text{dof}$ does not change much. However, raising $\omega_0/T$ further
will finally make the model not fit the data anymore.
For the electrical conductivity, we can include its maximal
deviation from the result obtained using the untruncated Ansatz as an upper 
systematical error, see 
Fig.~\ref{fig_spf_plot} (\textit{Bottom right}).
The corresponding 
spectral function with the cut applied at $\omega_0/T=3$ is shown in 
Fig.~\ref{fig_spf_plot} for all three temperatures.\\
In our standard Ansatz $\rho_{\text{ans}}$ we model
the large frequency 
behavior as a scaled free continuum spectral function. Another approach
would be to instead incorporate a (higher order) perturbative calculation of 
the vector channel spectral function.
We choose to follow the strategy adopted in 
\cite{Burnier2012} and utilize a spectral function that consists of 
two contributions. First, the leading order (LO) expression for the vector spectral 
function in a thermal environment, which just corresponds to 
$\rho_V^{\text{free}}(\omega)$.
Second, we multiply it with the R-ratio in the vacuum, computed to $\text{N}^4\text{LO}$ 
\cite{Baikov2008,Baikov2009}, altogether leading to 
\begin{eqnarray}
	\label{eqn_definition_perturbative}
	\rho_{\text{impr}}(\omega,T)\equiv\rho_V^{\text{free}}(\omega,T)R(\omega^2).
\end{eqnarray}
In this case we still incorporate a factor multiplying the perturbative spectral 
function, $C$, to account for modifications from 
the surrounding medium, uncertainties in the renormalization etc.
The modified Ansatz thus is given by 
\begin{eqnarray}
	\label{eqn_modify_perturbative}
	\rho_{\text{R}}(\omega,T)=\rho_{\text{BW}}(\omega,T)
	+ C\rho_{\text{impr}}(\omega,T).
\end{eqnarray}
\begin{table}[b]
	\begin{tabular}{|c||c|c|c|c|c|}
		\hline
		$T$ & $\sigma/(C_{\text{em}}T)$ & $\Gamma/T$ & $c_{BW}T/\Gamma$ & $C$ & $\chi^2/\text{dof}$ \\
		\hline
		$1.1T_c$ & $0.452(251)$ & $1.62(1.09)$ & $0.790(438)$ & $0.993(7)$ & $1.11$ \\
		$1.3T_c$ & $0.301(87)$ & $2.89(1.18)$ & $0.504(145)$ & $0.984(8)$ & $0.53$ \\ 
		$1.5T_c$ & $0.326(87)$ & $2.38(85)$ & $0.548(146)$ & $0.996(7)$ & $1.12$ \\
		\hline
	\end{tabular}
	\caption{Results of fitting the Ansatz $\rho_{\text{R}}$ for all three temperatures.}
	\label{tab_fit_results_pert}
\end{table}
Fitting our data with the Ansatz $\rho_{\text{R}}$ and listing the results 
in Tab.~\ref{tab_fit_results_pert}, we generally find that 
the transport peak becomes a bit narrower and higher, when compared to $\rho_{\text{ans}}$,
with the most pronounced
effect at $T=1.1T_c$, where the peak rises one third in height.
However, the strong effect at $1.1T_c$ is accompanied by huge errors
\begin{figure}[t]
	\centering
	\includegraphics[width=0.495\textwidth]{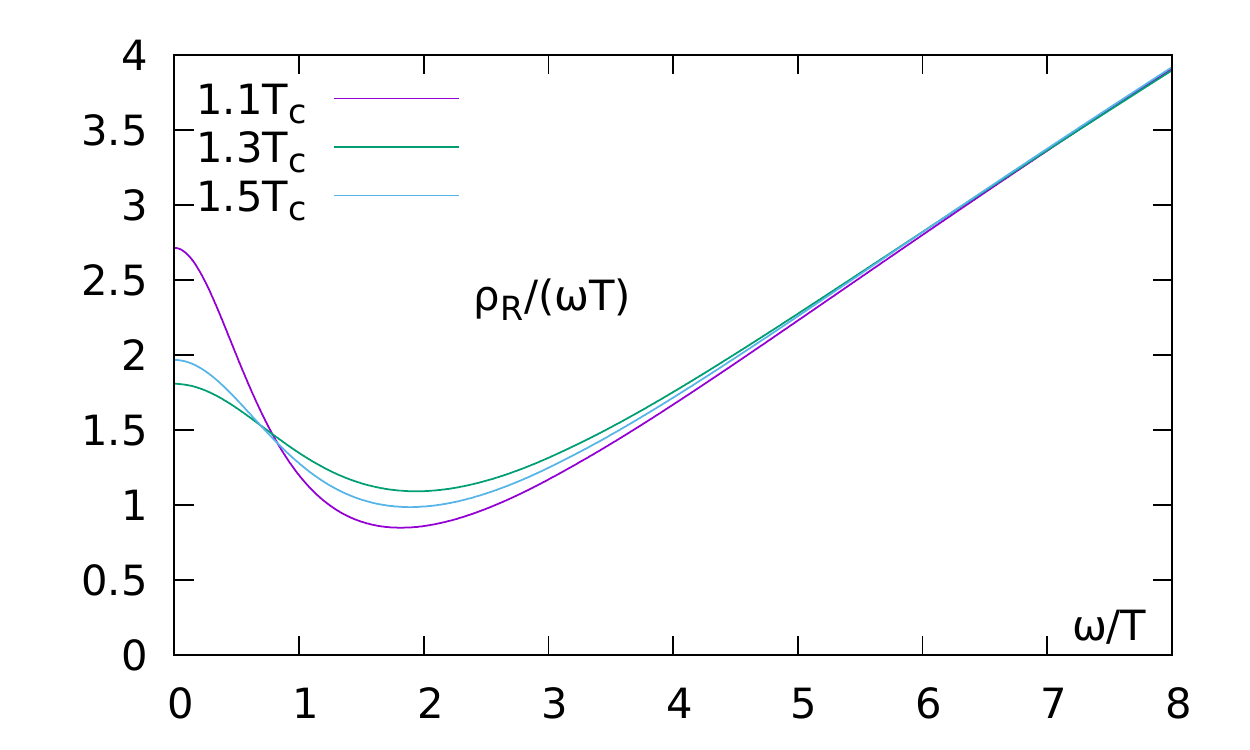}
	\caption{The resulting spectral function when utilizing perturbative input.}
	\label{fig_modify_perturbative}
\end{figure}
of both the transport peak's width and height, of $50-80\%$. The resulting 
spectral functions for all three temperatures 
are shown in Fig.~\ref{fig_modify_perturbative}.
\begin{figure*}[t]
	\centering
	\includegraphics[width=0.495\textwidth]{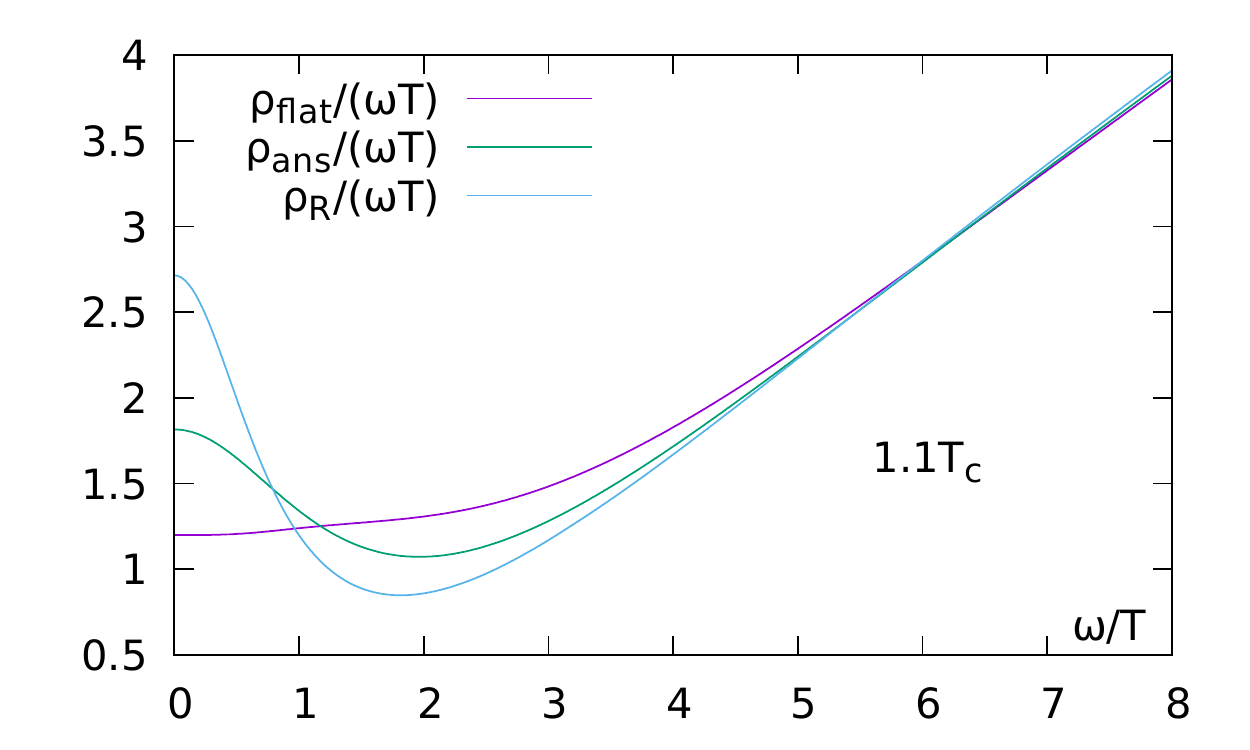}
	\includegraphics[width=0.495\textwidth]{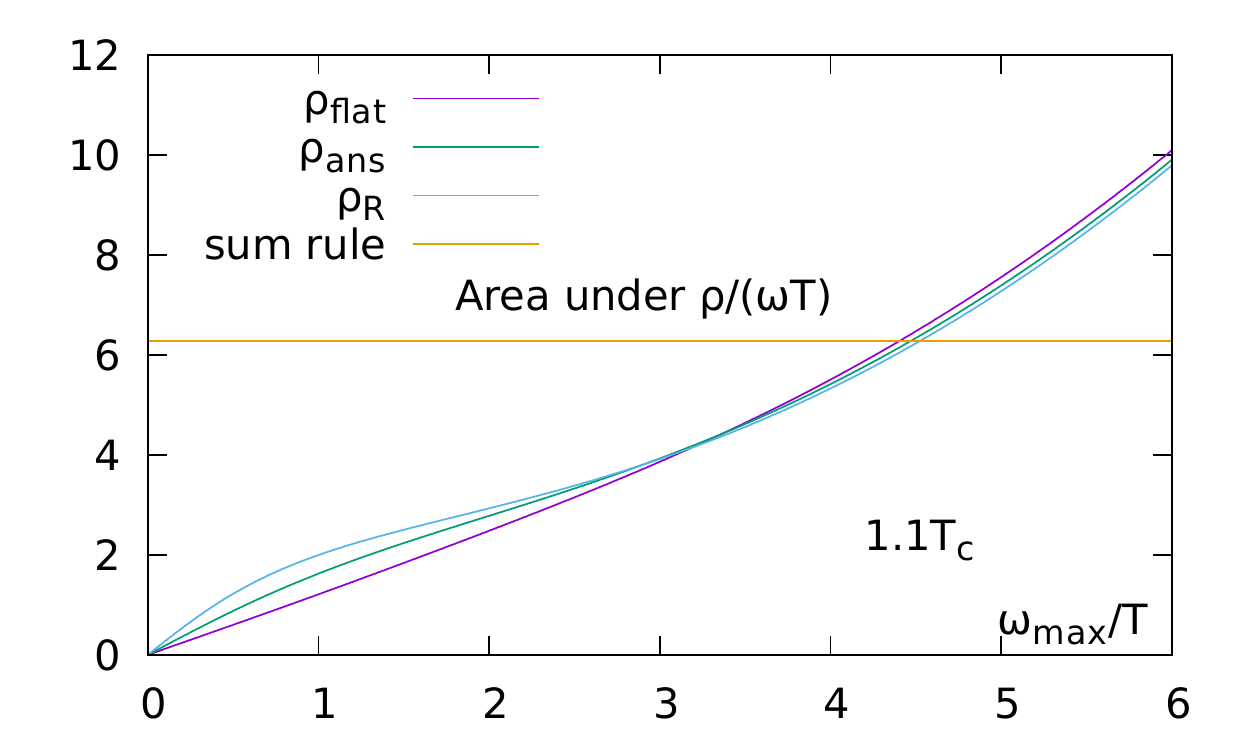}
	\includegraphics[width=0.495\textwidth]{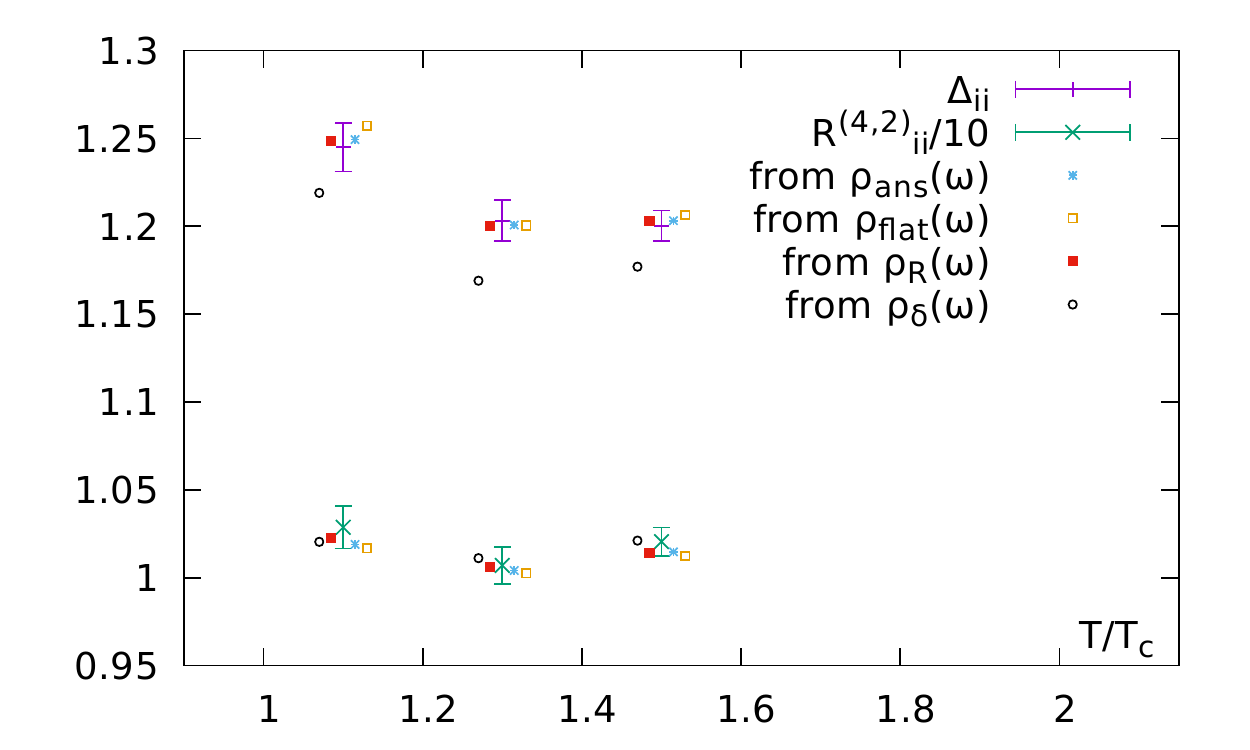}
	\caption{\textit{Top left}: The solutions of different Ans\"atze compared for $T=1.1T_c$. 
		Note that the difference between $\rho_{\text{R}}$ and $\rho_{\text{ans}}$
		is most pronounced at $T=1.1T_c$.
		\textit{Top right}: Integrating $\rho/(\omega T)$ up to $\omega_{\text{max}}$, i.e.
		numerically computing its primitive function for $T=1.1T_c$.
		\textit{Bottom}: The thermal moments for all $T$ and their respective reconstructions
		from the fit, shown for all three models $\rho_{\text{ans}}$,
	$\rho_{\text{flat}}$ and $\rho_{\text{R}}$.}
	\label{fig_compare_models}
\end{figure*}
The 
parameter $C$ is smaller than unity in all cases, and for $1.1T_c$ and 
$1.5T_c$ it is even compatible with unity within its errors.
From (\ref{eqn_modify_perturbative}) we see, comparing to the 
large frequency part of $\rho_{\text{ans}}(\omega)$, 
that the factor $(1+k)$ corresponds to a factor of $CR(\omega^2)$ 
in the improved case. On the one hand this makes the improvement 
of the large frequency part explicit, as the correction coefficient 
now depends on the frequency. On the other hand, from a purely technical 
point of view, the remaining \textit{correction constant} $C$ becomes less important
for the fit itself, as its deviation from unity is small, and partly 
negligible within its errors. 
To state a final result from this Ansatz, we plotted the maximum and minimum 
electrical conductivity, with errors coming from the fit, as the respective 
right bar of the paired bars 
in Fig.~\ref{fig_spf_plot}
(\textit{Bottom right}).\\

\section{Results}
Comparing the three models $\rho_{\text{ans}}$, $\rho_{\text{flat}}$
and $\rho_{\text{R}}$
in Fig.~\ref{fig_compare_models} (\textit{top left}), we see that the area under each peak of 
the spectral functions is very similar. From a rather sharp peak to a fully flat
behavior, all solutions are equally good ones in terms of 
stability and $\chi^2$. 
Fig.~\ref{fig_compare_models} (\textit{top right}) shows the primitive 
integral of $\rho/(\omega T)$ for all three cases, 
which reveals that there is a range of frequencies,
roughly $\omega/T\gtrsim3$, for which the areas under the curves are the same.
There is a sum rule found in perturbation theory \cite{Moore2006}, which states that 
the area under $\rho/\omega$ over the peak region is independent of the coupling, i.e.
fixed for our purposes. 
The authors compute an explicit expression in the framework of kinetic 
theory, given by 
\begin{eqnarray}
	\int \dd\br{\frac{\omega}{T}} \frac{\rho\br{\frac{\omega}{T}}}{\omega T}
	=\frac{2\pi}{3}N_c,
\end{eqnarray}
\begin{figure*}[t]
	\centering
	\includegraphics[width=0.495\textwidth]{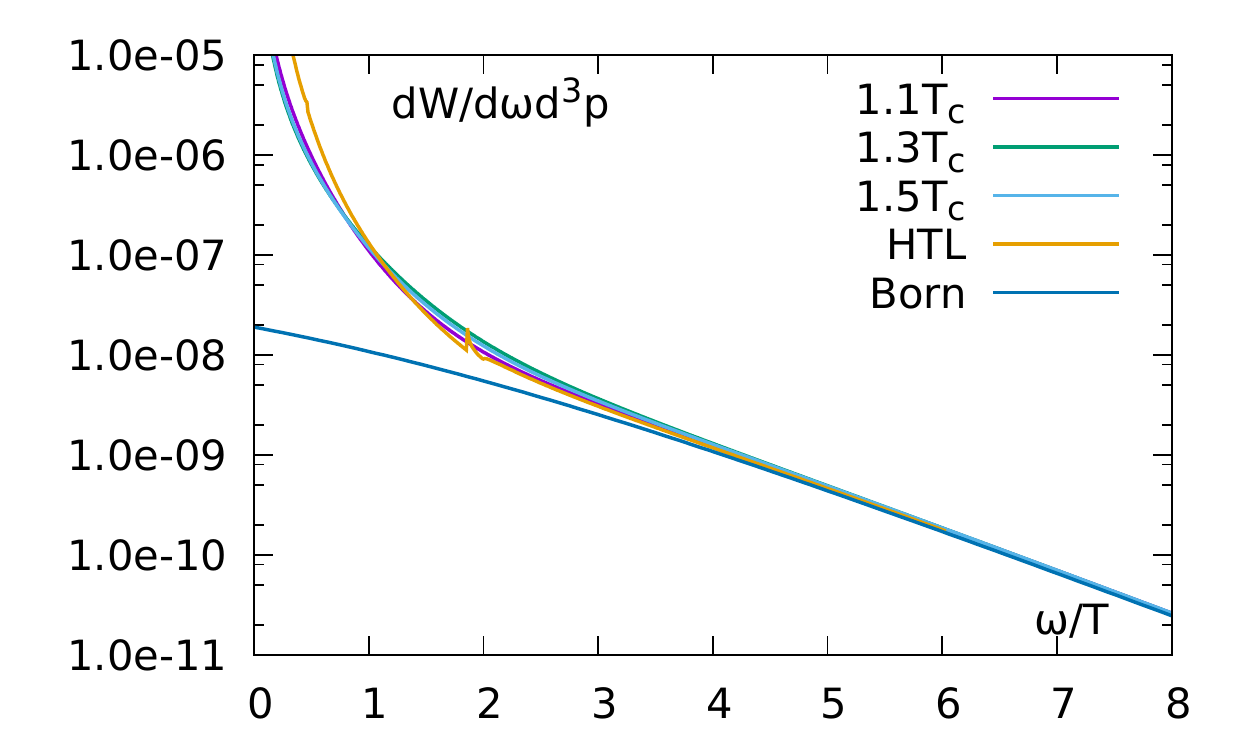}
	\includegraphics[width=0.495\textwidth]{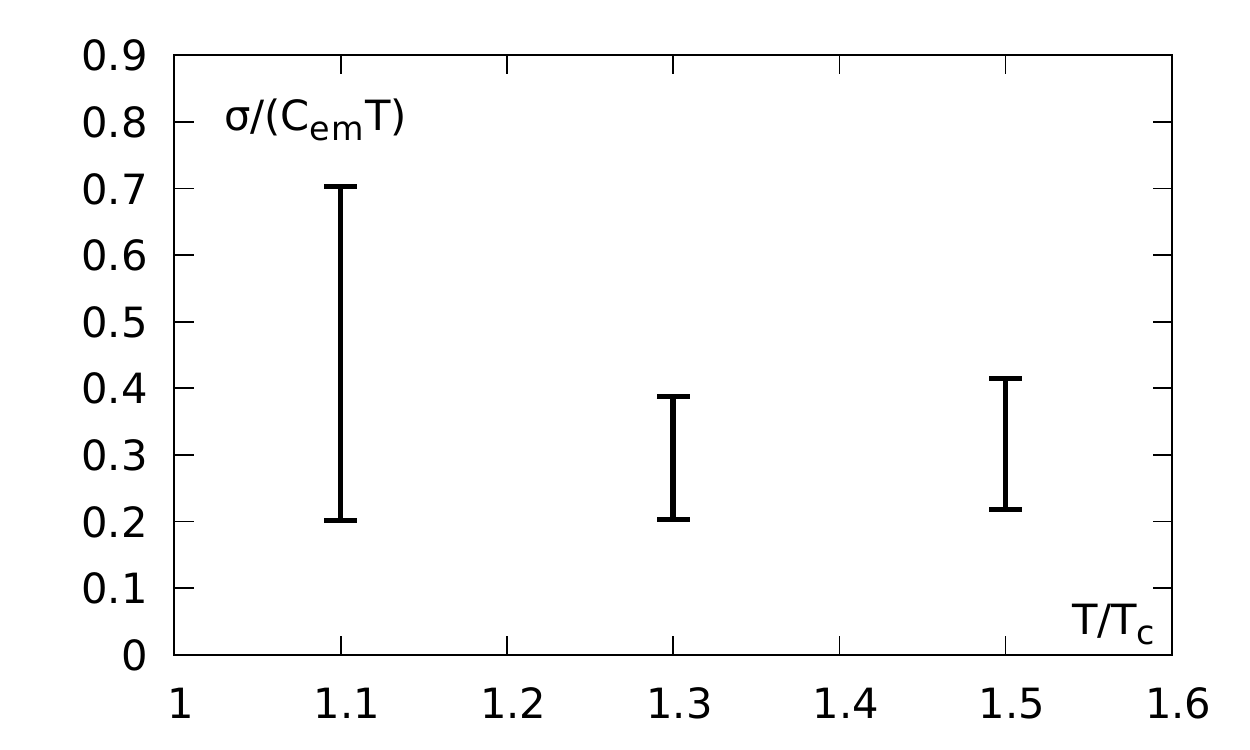}
	\caption{\textit{Left}: The thermal dilepton rate as obtained from $\rho_{\text{R}}$ 
		as a function of $\omega/T$, accompanied by the HTL rate and 
		the non interacting Born rate.
		\textit{Right}: The final results for the electrical conductivity. They incorporate 
		the full systematics, i.e. the minimum and maximum conductivities, 
	respectively, of $\rho_{\text{ans}}$ and $\rho_{\text{R}}$.}
	\label{fig_final_results}
\end{figure*}
where we suppressed a factor $C_{em}$, which is not contained in the spectral 
function we obtain from the continuum extrapolated correlators.
We plot this value as a straight line for reference.
Information about the exact shape of the spectral function is difficult
to obtain using Euclidean data, 
because for small frequencies $K(\omega,\tau,T)\to2/\omega$,
independent of $\tau$, and thus (\ref{eqn_integ_trans}) is fulfilled for any 
spectral function whose low frequency region has the correct area.
This effect we clearly also see in our fit procedure working on 
non-perturbative continuum data.
The sum rule also reflects the small electrical 
conductivities obtained by fitting $\rho_{\text{flat}}$, compared 
to the other Ans\"atze, that inhibit peaked structures.
As discussed in Sec.~\ref{sec_fitting}, the width of the peaks, 
in the Ans\"atze that feature a Breit-Wigner peak
at small frequency, is characteristic for the strength of the 
interactions in the medium.
We generally find $\Gamma/2\sim \Obs(T)$ in all of our fits. 
A width of this order is typical for a 
strongly coupled regime, and the corresponding scale for 
a weakly coupled plasma would be $\Gamma/2\sim\Obs(gT)$ \cite{Meyer2011}, which is 
parametrically smaller.
Thus our fitting results, obtained using the Ans\"atze motivated from
kinetic theory and from the AdS/CFT correspondence, are both in agreement with
expectations of a rather
strongly coupled medium from $1.1T_c$ to $1.5T_c$
and with qualitative features expected from perturbation theory,
although perturbative estimates usually overestimate the electrical
conductivity for example.
Calculations at finite momenta may offer an alternative way to estimate 
the electical conductivity as the system becomes more perturbative
at the relevant scales 
as observed in the study in \cite{Ghiglieri:2016tvj}, which 
can be used to further reduce the systematic uncertainties of the spectral function also
at zero momentum.\\
As mentioned before, generally the fits of the models $\rho_{\text{ans}}$, 
$\rho_{\text{part}}$ and $\rho_{\text{flat}}$ 
to the correlator data yield
equally good results, 
which shows the difficulties in resolving details of the
transport region of the spectral function. 
We find that utilizing the covariance of the data points in the fit 
generally increases the resolution of the procedure with respect to the low frequency 
region, as described in Sec.~\ref{sec_squeezing_BW}, and also generally enhances the 
quality of the fit, in the sense that the errors on the resulting parameters 
are smaller compared to fits without the covariance matrix. 
The role of the second thermal moment as a constraint in the fit, however, 
turns out to be a more subtle one: On the one hand, when ignoring the covariance of the data,
fitting with the second thermal moment as a constraint essentially also shows
the effect of reducing the errors on the resulting fit parameters, as opposed 
to not constraining the fit with the thermal moment. But this effect does 
not appear when fitting with the full covariance of the data, showing that 
the information on the curvature of the correlation function is
already largely contained in the statistical correlation. 
On the other hand, in the fit of $\rho_{\delta}(\omega)$, done with 
the full covariance of the data, it still serves as a very strong indication
that the fit breaks down.
This observation motivates us to 
also show the
reconstruction of the extrapolated second thermal moment and the ratio 
of fourth to second thermal moments in
Fig.~\ref{fig_compare_models} (\textit{bottom}) for all temperatures. 
The reconstructed values from the fits 
(apart from $\rho_{\delta}(\omega)$) 
generally are in accord with the second thermal moments as extracted from the data,
which underlines that our fits work well from the point of view of fit 
quality. Although the second thermal moment is especially sensitive to the 
low frequency region of the spectral function, at the current state of data accuracy
we cannot clearly differentiate between the models 
$\rho_{\text{ans}}$, $\rho_{\text{R}}$ and $\rho_{\text{flat}}$ 
using this observable. Considering that 
for $T=1.1T_c$ and $T=1.5T_c$ the thermal moment for $\rho_{\text{flat}}$ deviates
from the data visibly, but within errors, increasing the accuracy of the 
thermal moments data might provide a handle for this.
The ratios $R_{ii}^{(4,2)}$ are not included in the fit as a constraint, but 
a posteriori (re)constructed from the data and resulting fit parameters, respectively.
They compare within errors, although for $T=1.1T_c$ and $T=1.5T_c$ the results from the 
fit do not compare well. Note that the value from $\rho_{\delta}$ compares 
as well as any other reconstructed value, unlike in the case of the second moments
discussed above. As expected in Sec.~\ref{sec_fitting} from a rather qualitative 
argument, we thus see here explicitly that the ratios of fourth to second thermal 
moment are indeed far less sensitive to the low 
frequency region than the second thermal moments.\\
Our final results for the electrical conductivity 
for all three temperatures are summarized in Fig.~\ref{fig_final_results} 
(\textit{Right}).
In the plot we show the respective minimum and maximum value resulting 
from the two Ans\"atze $\rho_{\text{ans}}$ and $\rho_{\text{R}}$, to 
incorporate the full systematics found in our analysis,
\begin{align*}
	\left.\frac{\sigma}{C_{\mathrm{em}} T}\right|_{1.1T_c} = \ 0.201-0.703\\
	\left.\frac{\sigma}{C_{\mathrm{em}} T}\right|_{1.3T_c} = \ 0.203-0.388\\
	\left.\frac{\sigma}{C_{\mathrm{em}} T}\right|_{1.5T_c} = \ 0.218-0.413
\end{align*}
In this temperature region they are comparable to recent lattice QCD results
using dynamical fermions \cite{Amato2013,Aarts:2014nba,Brandt:2015aqk}. 
Note that in these studies a drop of the electrical conductivity is observed
when going to smaller temperatures around $T_c$, which may be due to the
different nature of the deconfinement transition. 
For a comparison of recent lattice QCD results see \cite{Brandt:2015aqk}
and a comparison of different determinations of the electrical conductivity can be found in 
\cite{Greif:2016skc}.\\
For a comparison of different calculations of the electrical conductivity
see \cite{Greif:2016skc}.
The resulting thermal dilepton rates, obtained from the spectral function
$\rho_{\text{R}}$
via the first expression of (\ref{eqn_dilrate}),
are shown in Fig.~\ref{fig_final_results} (\textit{left}) for all three temperatures 
and a sum of squared charges of $C_{em}=\sum_i q_i^2=5/9$, corresponding to 
two valence quark flavors $u$ and $d$. 
Our rates are qualitatively comparable to the rate obtained by an HTL calculation
\cite{Braaten1990} in the large frequency region, as well as to the 
leading order (Born) rate. However, compared to the HTL computation, 
our results show an enhancement in the intermediate region $\omega/T\sim2$ and a 
qualitatively different behavior for small frequency, as the leading term for 
$\omega\to0$ is different (see also Fig.~\ref{fig_spf_plot}).
Finally, the soft photon rate is 
given for all temperatures by the electrical conductivity via 
(\ref{eqn_photonrate_conductivity}), and $C_{em}=5/9$, as 
\begin{align*}
	\left.\omega\frac{\dd R_{\gamma}}{\dd p^3}\right|_{1.1T_c}&=&\bc{5.00-17.48}\times10^{-5}T_c^2,\\
	\left.\omega\frac{\dd R_{\gamma}}{\dd p^3}\right|_{1.3T_c}&=&\bc{7.05-13.47}\times10^{-5}T_c^2,\\
	\left.\omega\frac{\dd R_{\gamma}}{\dd p^3}\right|_{1.5T_c}&=&\bc{10.08-19.18}\times10^{-5}T_c^2.
\end{align*}
The photon rates at the two higher temperatures show a slight trend to
rise with temperature, 
but this is within errors, and for the lower bound alone this trend is 
true for all $T$. However, the lowest temperature suffers from a large 
upper bound, that is also seen in the determined electrical conductivity.

\section{Conclusion and outlook}
Using non-perturbatively improved Wilson Clover valence fermions we 
performed continuum extrapolations of light vector channel correlation
functions at three temperatures. The extrapolations yield reliable results
with errors at the sub-percent level. A consequence of bootstrapping the 
extrapolation is that the covariance matrix of the data can be computed and 
is shown to permit stable fits.
Employing a phenomenologically motivated Ansatz for the corresponding 
spectral function, these are used to perform
a fully correlated $\chi^2$-minimization and to
obtain results for the spectral functions and thus the electrical 
conductivities via a Kubo relation, the thermal dilepton rates and the 
soft photon rates. 
The second thermal moments, obtained from a separate continuum extrapolation, are 
found to be sensitive to the low frequency region of the spectral function, while 
the ratios of the fourth to the second thermal moment are sensitive to a region at larger 
frequency.
Different systematics related to the Ansatz are investigated. We find 
an essential improvement of the fit with respect to the low frequency region 
when performing the fit fully correlated, as opposed to neglecting the covariances 
of the data. 
Fitting a form of Ansatz inspired by the phenomenology of 
a strongly coupled QGP shows a comparable fit quality to the Ansatz motivated 
by a quasiparticle description,
which implies that our procedure 
at this time does not resolve a difference between these two differently shaped 
spectral functions.
This difficulty is reflected by the fact that the different spectral functions,
extracted from our non-perturbative data, all 
fulfill a sum-rule that is valid in the low frequency region.
However, by observing 
the resulting peak widths from the fits of a Breit-Wigner peak, we find that 
they
are of the order of $\Gamma/2\sim \Obs(T)$
which reveals that both the peaked Ans\"atze and the flat Ansatz hint at 
a strongly coupled medium.
The use of a perturbative estimate for the large frequency part of the 
spectral function is found to generally increase the upper bound of the 
electrical conductivity. 
The electrical conductivities are in accordance 
with earlier results obtained by MEM and $\chi^2$-minimization methods. We find 
no significant temperature dependence in the temperature range investigated, as was
expected from the weak temperature dependence of the correlation functions.
The thermal dilepton rates are compared to the HTL and leading
order rates and show almost no temperature dependence in the analyzed temperature 
region, either. The lower bound on the determined soft photon rates clearly 
follows a trend by rising with temperature. However, the overall large errors,
especially at $T=1.1T_c$, make it difficult to determine a general trend.\\
The use of a higher order perturbative estimate for the large frequency behavior 
of our Ansatz opens two concrete possibilities. First, 
because for two temperatures the resulting the constant in front of the large
frequency part of the spectral function is compatible with unity within errors, 
we mark that in this sense 
further improvements might make it superfluous and thus reduce the number 
of parameters in the fit. Second, the low frequency behavior of the perturbative 
estimate is merely leading order. By incorporating additional, possibly perturbative, 
input there, the resolution of the fit in the low frequency region might increase.
A fit with a general polynomial Ansatz for the low frequency region, and 
constraints on a smooth connection to the perturbative large frequency behavior, 
has been done in \cite{Ghiglieri:2016tvj} on the same data as used in this
work leading to results in agreement with the ones presented here. 
Additionally, fits were performed to the same continuum extrapolated vector correlators,
but featuring non-vanishing spatial momenta, which allows for an evaluation of photon rates at frequencies
$\omega>0$ and also opens the possibility to estimate the electrical
conductivity and diffusion coefficients in general from correlation functions
at non zero momenta.
A natural extension of these studies should be an investigation at lower
temperatures closer to $T_c$ and below. As light quark degrees of freedom will become
more important and vector resonance contributions emerge at lower temperatures,
this is well justified only in the presence of dynamical quarks.

\begin{acknowledgments}
	The results have been achieved using the PRACE Research Infrastructure
	resource JUGENE based at the J\"ulich Supercomputing Centre in Germany,
	the OCuLUS Cluster at the Paderborn Center for Parallel Computing
	and the
	Bielefeld GPU-cluster resources. 
	This work has been partly supported by BMBF under grants 05P12PBCTA and
	56268409, NSFC under grant no. 11535012 and the GSI BILAER grant. 
	We would like to 
	thank M.~Laine and J.~Ghiglieri and the members of the
	Bielefeld-BNL-CCNU collaboration for fruitful discussions, as well as 
	A.~Francis and M.~M\"uller for collaboration at initial stages of
	this project.
\end{acknowledgments}

\bibliography{paper}

\end{document}